\documentclass{ws-ijmpa}
\usepackage[super,compress]{cite}
\usepackage{graphicx}
\begin{document}
\markboth{N. Yamanaka}{Review of the EDM of light nuclei}

%
\catchline{}{}{}{}{}
%

\title{Review of the electric dipole moment of light nuclei
}
\def\vc#1{\mbox{\boldmath $#1$}}

\author{Nodoka Yamanaka
}

\address{
iTHES Research Group, Riken, 2-1 Hirosawa\\
Wako, 351-0198 Saitama, Japan\\
nodoka.yamanaka@riken.jp\\
and\\
Complex Simulation Group, School of Biomedicine,\\
Far Eastern Federal University,\\
Vladivostok, 690950 Russia\\
}



\maketitle


\begin{abstract}
In this review, we summarize the theoretical development on the electric dipole moment of light nuclei.
We first describe the nucleon level CP violation and its parametrization.
We then present the results of calculations of the EDM of light nuclei in the ab initio approach and in the cluster model.
The analysis of the effect of several models beyond standard model is presented, together with the prospects for its discovery.
The advantage of the electric dipole moment of light nuclei is focused in the point of view of the many-body physics.
The evaluations of the nuclear electric dipole moment generated by the $\theta$-term and by the CP phase of the Cabibbo-Kobayashi-Maskawa matrix are also reviewed.


\end{abstract}

\ccode{PACS numbers:11.30.Er,21.10.Ky,24.80.+y}



\section{Introduction\label{sec:intro}}

Our Universe is filled of matter, and no macroscopic quantity of antimatter can be found.
The asymmetry between the number of matters against that of antimatters is expressed in terms of the matter-photon ratio $\eta = \frac{n_B}{n_\gamma}$, where the number of photons is the synonym of that of the pairs of a particle and its antiparticle which have populated the early Universe, and recent observation of Planck is giving $\eta = O(10^{-10})$ \cite{planck}.

To generate the baryon number asymmetry in our Universe, several conditions have to be satisfied.
Those criteria were first given by Sakharov, and the requirements are (1) the existence of baryon number violation, (2) the violation of the discrete $C$ and $CP$ symmetries, and (3) the departure from thermal equilibrium \cite{sakharov}.
Attempts to explain the baryon number asymmetry within the standard model (SM) have been made \cite{farrar,huet}, but it is now known that the CP violation brought by the Cabibbo-Kobayashi-Maskawa (CKM) matrix \cite{ckm,jarlskog} is in great deficit.
The existence of matters around us is thus suggesting the existence of new physics beyond standard model (BSM), and the search for new CP violating sources beyond it is now one of the most important subjects of particle physics.

The search for CP violation, since its first observation in $K$ meson decays \cite{cronin}, was performed in many systems.
One of the most attractive observables in this context is the {\it electric dipole moment} (EDM) \cite{hereview,bernreuther,khriplovichbook,ginges,pospelovreview,fukuyamareview,hewett,engelreview,yamanakabook,devriesreview}.
The search for EDMs has several notable advantages.
The first one is its accurate experimental measurability.
It does not involve final state interaction effects, which are the main background mimicking the CP-odd observables in decays or reactions of particles.
The CKM contribution to the EDM is also known to be very small from several analyses.
Those two facts greatly reduce the source of systematics of EDM experiments, and help us to improve the experimental accuracy beyond that of accelerator based experiments.
The second advantage is that the EDM can be measured in various systems, and that composite systems may enhance the fundamental level CP violation through the nontrivial many-body effects.
The third strong point is that in many cases, the cost to prepare experiments are cheaper than other approaches, such as the accelerator based ones.
Owing to those advantages, the EDM was extensively studied in experiments.

The principle of the measurement of the EDM is to observe the spin precession under an electric field \cite{purcell}.
The EDM was so far measured in charge neutral systems, such as the neutron \cite{baker}, atoms and molecules\cite{rosenberry,regan,griffith,hudson,acme,parker,graner,bishof}, for which the spin precession frequency can be probed with high accuracy by applying simultaneously magnetic and electric fields.
For the measurement of the EDM of charged systems, the same method does not work due to the acceleration by the electric field, but it is possible to apply an effective electric field to them by rotating them under controlled magnetic and electric fields \cite{muong2,storage1,storage2,storage3,storage4,storage5,storage6,storage7}.
Recently, new techniques to measure the EDM of charged particles using storage rings are becoming available \cite{storage8,storage9,storage10,bnl}, and the measurement of the EDM of light nuclei is receiving much attention.
The measurement of the nuclear EDM has several notable advantages.
First, Schiff's screening phenomenon \cite{schiff}, encountered in the atomic systems, is not relevant for a bare nuclear system, and important suppression of the nuclear level CP violation is avoided.
The second strong point is the prospective experimental sensitivity of $d_A \sim O(10^{-29})e$ cm, which is well below the current experimental lower limit of the neutron EDM $d_n < 2.9 \times 10^{-26}e$ cm \cite{baker}.
Finally, we have the possibility of a nuclear level enhancement of the CP violation due to the many-body effect \cite{sushkov,sushkov2,sushkov3,ginges}.

To discover new sources of CP violation, the realization of experiments sensitive to them are required.
For that, finding nuclear systems where the CP violation is enhanced is essential.
An important work is therefore to evaluate with a reasonable accuracy the nuclear many-body effect and theoretically find nuclei with large enhancement.
We must also theoretically control the nuclear EDM of various nuclei to disentangle the contribution of the new physics which lies in a very wide parameter space.
In this review we present recent developments on the theoretical investigations of the EDM of light nuclei in these directions.

This review is organized as follows.
We first define the nuclear EDM and introduce the nucleon level CP violating interaction.
We then show and analyze in Section \ref{sec:intrinsicnucleonedm} the results of the calculation of the intrinsic nucleon EDM contribution to the nuclear EDM.
In Section \ref{sec:abinitio}, we present the result of ab initio investigations of the EDM of the deuteron, $^3$H, and $^3$He.
In Section \ref{sec:cluster}, we discuss the theoretical evaluation of the EDM of light nuclei in the cluster model.
We then present the Strong CP problem caused by the $\theta$-term and its effect on the nuclear EDM as the uncertainty of SM.
To ensure that the SM contribution due to the CKM matrix is a negligible background, we discuss it in Section \ref{sec:ckmedm}.
We then analyze in Section \ref{sec:prospects} the constraints on new physics BSM and future prospects on the study of the nuclear EDM.
Finally, we summarize the discussion.

\section{Nuclear electric dipole moment and CP-odd hamiltonian}

\subsection{Electric dipole moment of nuclear many-body systems}

\begin{figure}[htb]
\begin{center}
\includegraphics[width=12cm]{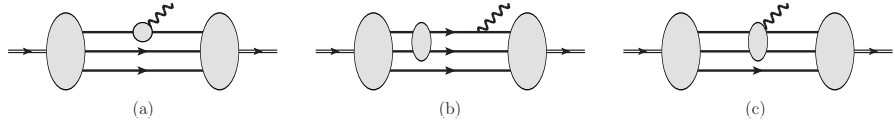}
\caption{\label{fig:nuclear_edm_breakdown}
Diagrammatic representation of the breakdown of the nuclear EDM: 
(a) Intrinsic nucleon EDM contribution, (b) Nuclear polarization due to the CP-odd nuclear force, (c) Exchange current effect.
Single and double lines denote the nucleon and nucleus, respectively.
The wavy lines are the electromagnetic current insertion which probes the EDM.
}
\end{center}
\end{figure}

There are three leading nucleon level CP violating processes which contribute to the nuclear EDM:
\begin{itemize}
\item
The intrinsic nucleon EDM.

\item
The nuclear polarization due to the CP-odd nuclear force.

\item
The exchange current effect.

\end{itemize}
The schematic picture of the breakdown of the nuclear EDM is shown in Fig. \ref{fig:nuclear_edm_breakdown}.

We first define the contribution of the intrinsic nucleon EDM to the nuclear EDM.
The nucleon EDM is proportional to the nucleon spin, and its effect is given by
\begin{eqnarray}
d_A^{\rm (Nedm)} 
&=&
\sum_{i=1}^A
d_i \langle \,A\, |\, \sigma_{iz} \, |\, A\, \rangle
\equiv 
\langle \sigma_p \rangle_A \, d_p + \langle \sigma_n \rangle_A \, d_n
,
\label{eq:intrinsicnucleonedm}
\end{eqnarray}
where $|\, A\, \rangle$ is the polarized nuclear wave function (in the $z$-axis). 
The proton and neutron EDMs are denoted by $d_p$ and $d_n$, respectively.
The nucleon spin matrix elements $\langle \sigma_p \rangle_A$ and $\langle \sigma_n \rangle_A$ can be calculated in nuclear physics, and they only depend on the nuclear structure.
In this review, they are also called ``enhancement factors'' or ``spin quenching factors''.

The next effect is the nuclear polarization generated by the CP-odd nuclear force.
This contribution is defined by
\begin{eqnarray}
d_{A}^{\rm (pol)} 
&=&
\sum_{i=1}^{A} \frac{e}{2} 
\langle \, \tilde A \, |\, (1+\tau_i^z ) \, z_{i} \, | \, \tilde A \, \rangle
,
\end{eqnarray}
where $|\, \tilde A\, \rangle$ is the polarized nuclear wave function.
The coordinate in the center of mass frame and the isospin Pauli matrix of the $i$th nucleon are denoted by $r_i$ and $\tau_i^z$, respectively.
This nuclear polarization is induced only if there is a mixing of parity and CP in the nuclear wave function $|\, \tilde A\, \rangle$.

Finally, the exchange current contribution is given by
\begin{eqnarray}
d_{A}^{\rm (ex)} 
&=&
\sum_{i < j =1}^{A}
\langle \, \tilde A \, |\, \int \rho^{(2)}_{ij} z_i d^3 \vc{r} \, | \, \tilde A \, \rangle
,
\end{eqnarray}
where $\rho^{(2)}_{ij}$ is the two-body exchange current operator acting on $i$th and $j$th nucleons.
The three-body exchange current operator and so forth can also be defined in a similar manner.
The exchange current is composed of the CP-even and CP-odd parts.
This effect is suppressed by the power of the nucleon velocity $(v/c)^2$, so we neglect it in this review.

\subsection{Nucleon level CP violation\label{sec:nucleonlevelcpv}}

The nuclear EDM is generated only if there is nucleon level CP violation.
The leading order nucleon level CP violation receives contribution from the following P, CP-odd chiral lagrangian \cite{chiral3nucleon,mereghetti2}:
\begin{eqnarray}
{\cal L}
&=&
-2 \bar N (\bar d_0 + \bar d_1 \tau^z) S^\mu N v^\nu F_{\mu \nu}
-\frac{1}{2 f_\pi} \bar N (\bar g_0 \vc{\tau} \cdot \vc{\pi} + \bar g_1 \pi^z ) N
\nonumber\\
&&
+m_N \Delta_{3\pi} \, \pi^z \vc{\pi}^2
+ \bar C_1 \bar N N \partial_\mu (\bar N S^\mu N )
+ \bar C_2 \bar N \vc{\tau} N \cdot \partial_\mu (\bar N S^\mu \vc{\tau} N )
,
\label{eq:cpv_chiral_lagrangian}
\end{eqnarray}
where $S^\mu = (0, \vc{\sigma}/2)$ and $v_\mu = (1, \vc{0})$.
The pion decay constant is given by $f_\pi = 93$ MeV and the nucleon mass by $m_N = 939$ MeV.

\begin{figure}[htb]
\begin{center}
\includegraphics[width=10cm]{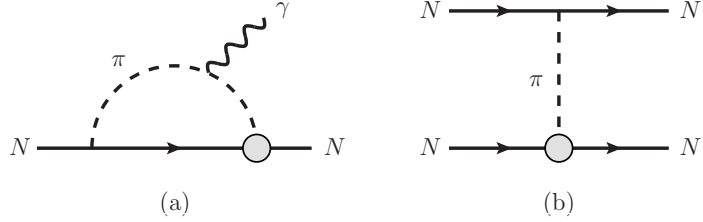}
\caption{\label{fig:cpv_pion-nucleon}
Leading contributions of the CP-odd pion-nucleon interaction to the nucleon level CP violation.
(a) Pion loop contribution to the nucleon EDM. (b) Pion exchange CP-odd nuclear force.}
\end{center}
\end{figure}

The intrinsic nucleon EDM receives the leading contribution from the bare nucleon EDM terms of the chiral effective lagrangian (\ref{eq:cpv_chiral_lagrangian}), and also from the pion loop diagram [see Fig. \ref{fig:cpv_pion-nucleon} (a)].
This radiative pion cloud effect is known to enhance the nucleon EDM \cite{crewther,pich,borasoy,narison}.
The isoscalar and isovector nucleon EDMs are given by \cite{crewther,ottnad,mereghetti1,devries1,guo1}
\begin{equation}
d_0
=
\bar d_0
+ \frac{e g_A \bar g_0 }{4 \pi f_\pi^2} \Biggl(
\frac{3 m_\pi}{4 m_N} 
\Biggr)
+ \frac{e g_A \bar g_1 }{16 \pi f_\pi^2} \frac{m_\pi}{m_N}
,
\label{eq:d_0}
\end{equation}
and 
\begin{equation}
d_1
=
\bar d_1 
+ \frac{e g_A \bar g_0 }{4 \pi^2 f_\pi^2} \Biggl(
\frac{2}{4-d} - \gamma_E+ \ln \frac{4 \pi \mu^2}{m_\pi^2}
+\frac{5 \pi m_\pi}{4 m_N} 
\Biggr)
+ \frac{e g_A \bar g_1 }{16 \pi f_\pi^2} \frac{m_\pi}{m_N}
,
\label{eq:d_1}
\end{equation}
respectively, where $g_A =1.27$ is the nucleon axial coupling \cite{ucna} and $m_\pi = 138$ MeV the pion mass.
Here we have neglected the isospin breaking terms of the pion and nucleon masses.
The bare nucleon EDMs $\bar d_0$ and $\bar d_1$ act as counterterms and cancel the divergence of the pion loop diagram.
The renormalization point $\mu$ is set to the hadronic scale, often between 1 GeV and 500 MeV.
At $\mu = 500$ MeV, all mesons except pions are integrated out.
We consider that the contributions from radiative processes involving $K$ mesons and hyperons \cite{fuyuto} are renormalized into $\bar d_0$ and $\bar d_1$.

The most important nucleon level CP violating process is the pion-exchange CP-odd nuclear force [see Fig. \ref{fig:cpv_pion-nucleon} (b)].
At the leading order, it is generated by combining the CP-odd pion-nucleon interaction with the CP-even one.
By taking the on-mass shell approximation of the nucleons, the momentum exchange representation of the leading CP-odd nuclear force in the chiral EFT yields \cite{maekawa,chiral3nucleon,mereghetti2}
\begin{eqnarray}
V_{P\hspace{-.5em}/\, T\hspace{-.5em}/\, } (\vc{k})
& = &
\bigg\{ 
i\frac{g_A \bar g_0}{f_\pi^2 (\vc{k}^2+m_\pi^2)}
({\vc{\tau}}_{1}\cdot {\vc{\tau}}_{2})\, {\vc{\sigma}}_{-}
+i\frac{g_A \bar g_1}{2 f_\pi^2 (\vc{k}^2+m_\pi^2)}
( \tau_{+}^{z}\, {\vc{\sigma}}_{-} +\tau_{-}^{z}\,{\vc{\sigma}}_{+} )
\nonumber\\
&& \ \ 
-\frac{i}{2} \Bigl[
\bar C_1 + \bar C_2 ({\vc{\tau}}_{1}\cdot {\vc{\tau}}_{2})
\Bigr]\, {\vc{\sigma}}_{-}
\bigg\}
\cdot
\hat{ \vc{k}} \,
,
\label{eq:CPVEFTpot}
\end{eqnarray}
where the subscripts label the nucleon.
The unit vector of the exchanged momentum $\vc{k}$ is given by $\hat{\vc{k}} \equiv \frac{\vc{k}_1 - \vc{k}_2}{|\vc{k}_1 - \vc{k}_2|}$.
The spin and isospin Pauli matrices are given by ${\vc{\sigma}}_{-} \equiv {\vc{\sigma}}_1 -{\vc{\sigma}}_2$, ${\vc{\sigma}}_{+} \equiv {\vc{\sigma}}_1 + {\vc{\sigma}}_2$, ${\vc{\tau}}_{-} \equiv {\vc{\tau}}_1 -{\vc{\tau}}_2$, and ${\vc{\tau}}_{+} \equiv {\vc{\tau}}_1 + {\vc{\tau}}_2$.
Here the contact couplings $\bar C_1$ and $\bar C_2$ contain the effects from short range physics, including those from heavy mesons with masses above the cutoff $\mu$.

\begin{figure}[htb]
\begin{center}
\includegraphics[width=10cm]{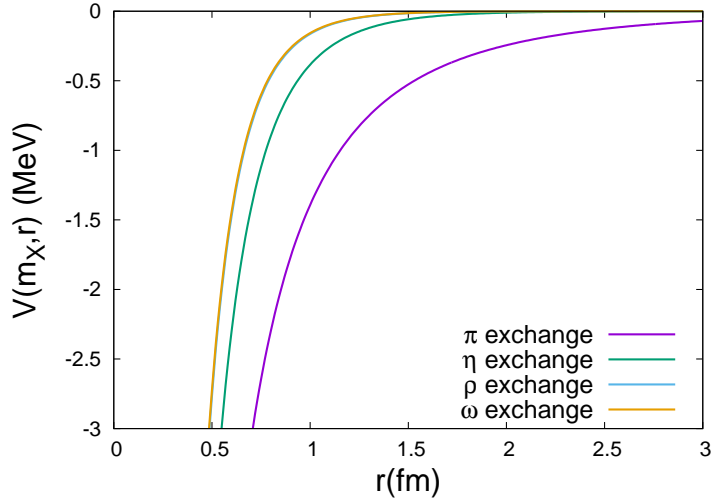}
\caption{\label{fig:cpvnn_meson}
The radial shape of the CP-odd nuclear force $V(m_X , r)$ for $\pi$-, $\eta$-, $\rho$- and $\omega$-exchanges.
}
\end{center}
\end{figure}

A more phenomenological CP-odd nuclear force based on one meson exchange model can be conceived \cite{Barton,pvcpvhamiltonian1,pvcpvhamiltonian2,pvcpvhamiltonian3}.
In the coordinate representation, it is given as
\begin{eqnarray}
V_{P\hspace{-.5em}/\, T\hspace{-.5em}/\, } (\vc{r})
& = &
\bigg\{ 
\bar{G}_{\pi}^{(0)}\, ({\vc{\tau}}_{1}\cdot {\vc{\tau}}_{2})\, {\vc{\sigma}}_{-}
+\frac{1}{2} \bar{G}_{\pi}^{(1)}\,
( \tau_{+}^{z}\, {\vc{\sigma}}_{-} +\tau_{-}^{z}\,{\vc{\sigma}}_{+} )
\nonumber\\
&&\hspace{11em}
+\bar{G}_{\pi}^{(2)}\, (3\tau_{1}^{z}\tau_{2}^{z}- {\vc{\tau}}_{1}\cdot {\vc{\tau}}_{2})\,{\vc{\sigma}}_{-} 
\bigg\}
\cdot
\hat{ \vc{r}} \,
V(m_\pi , r)
\nonumber\\
&&
+ \bigg\{ 
\bar{G}_{\eta}^{(0)}\, {\vc{\sigma}}_{-}
+\frac{1}{2} \bar{G}_{\eta}^{(1)}\,
( -\tau_{+}^{z}\, {\vc{\sigma}}_{-} +\tau_{-}^{z}\,{\vc{\sigma}}_{+} )
\bigg\}
\cdot
\hat{ \vc{r}} \,
V(m_\eta , r)
\nonumber\\
&& +
\Biggl\{
-\bar{G}_{\rho}^{(0)}\, ({\vc{\tau}}_{1}\cdot {\vc{\tau}}_{2})\, {\vc{\sigma}}_{-}
-\frac{1}{2} \bar{G}_{\rho}^{(1)}\,
( \tau_{+}^{z}\, {\vc{\sigma}}_{-} -\tau_{-}^{z}\,{\vc{\sigma}}_{+} )
\nonumber\\
&&\hspace{11em}
-\bar{G}_{\rho}^{(2)}\, (3\tau_{1}^{z}\tau_{2}^{z}- {\vc{\tau}}_{1}\cdot {\vc{\tau}}_{2})\,{\vc{\sigma}}_{-} 
\bigg\}
\cdot
\hat{ \vc{r}} \,
V(m_\rho , r)
\nonumber\\
&& +
\Biggl\{
- \bar{G}_{\omega}^{(0)}\, {\vc{\sigma}}_{-}
-\frac{1}{2} \bar{G}_{\omega}^{(1)}\,
( \tau_{+}^{z}\, {\vc{\sigma}}_{-} +\tau_{-}^{z}\,{\vc{\sigma}}_{+} )
\bigg\}
\cdot
\hat{ \vc{r}} \,
V(m_\omega , r)
,
\label{eq:CPVhamiltonian}
\end{eqnarray}
where the radial function $V$ is given by
\begin{equation}
V(m_X,r)
= 
\frac{1}{2m_N}
\vc{\nabla} \frac{e^{-m_X r }}{4 \pi r}
=
-\frac{m_X}{8\pi m_N} \frac{e^{-m_X r }}{r} \left( 1+ \frac{1}{m_X r} \right)
\ .
\label{eq:barecpvnn}
\end{equation}
The shape of $V(m_X , r)$ is shown in Fig. \ref{fig:cpvnn_meson} for the exchanges of $X = \pi$, $\eta$, $\rho$, $\omega$ mesons.
This CP-odd potential has been used in many previous ab initio analyses \cite{korkin,liu,stetcu,afnan,song,yamanakanuclearedm} and has served as a benchmark of the ab initio nuclear EDM calculations.

Let us here mention several features of the CP-odd nuclear force (\ref{eq:CPVhamiltonian}) and their expected impact on the nuclear polarization.
The CP-odd nuclear force has the structure of a product between the nucleon density and the three dimensional divergence of the spin density.
As it is a spin dependent interaction, nuclei with closed spin shell cannot be polarized.
The polarization is also dependent on the nucleon density, so that the increase of the nucleon number should enhance the nuclear EDM.
The CP-odd nuclear force is however an exchange of massive mesons, so its effect is limited in a finite range.
As the nuclear density is saturated and the nuclear volume increases for large nuclei, the CP-odd interaction will know a limitation of the nuclear polarization when the nucleon number sufficiently grows.
Another important feature is the derivative, which makes the CP-odd interaction to be sensitive on the surface of the nucleus.
This property suggests that nuclei with a well developed cluster structure may enhance the EDM, due to the increase of the density gradient inside the system.
The nuclear EDM may also be enhanced due to octupole deformations, which facilitate the transition between opposite parity states by decreasing their level spacings in the spectrum \cite{auerbach,Spevak,dobaczewski}.

The P, CP-odd $NN$ couplings $\bar G_X^{(i)}$ $(i=0,1,2 ; X = \pi , \eta , \rho , \omega)$ are dimensionless.
The isoscalar and isovector couplings can be given from the chiral EFT analysis.
To the leading order, the CP-odd pion-nucleon couplings and the CP-odd $NN$ couplings are related by
\begin{eqnarray}
\bar G^{(0)}_\pi 
&=& 
\frac{g_A m_N}{2 f_\pi^2} \bar g_0
,
\label{eq:g0pi}
\\
\bar G^{(1)}_\pi
&=& 
\frac{g_A m_N}{2 f_\pi^2} \bar g_1
.
\label{eq:g1pi}
\end{eqnarray}
In the chiral analysis, the isotensor CP-odd nuclear force [term with $\bar G^{(2)}_\pi$ in Eq. (\ref{eq:CPVhamiltonian})] is suppressed by an additional factor of light quark mass, so it is negligible in the analysis of the EDM of light nuclei.
(For heavy nuclei, it may be relevant due to the effective interaction generated by the isospin violation of the nuclear medium).

The chiral EFT analysis does not include mesons heavier than the pion, due to the cutoff placed near $\mu =$ 500 MeV.
Instead, their effect is renormalized into the contact interaction [terms with $\bar C_1, \bar C_2$ in Eq. (\ref{eq:cpv_chiral_lagrangian})].
In the hamiltonian (\ref{eq:CPVhamiltonian}), we can find terms with the same spin and isospin structures as the contact terms.
As the cutoff is lower than the masses of $\eta$, $\rho$ and $\omega$ mesons, their exchange processes can be approximated as contact interactions.
We can therefore match the contact couplings $\bar C_1$ and $\bar C_2$ with those of $\eta$ and $\rho$ exchange, respectively, as
\begin{eqnarray}
\bar G_\eta^{(0)}
&\approx &
-2m_N m_\eta^2 \bar C_1
,
\label{eq:contact1eta}
\\
\bar G_\rho^{(0)}
&\approx &
2m_N m_\rho^2 \bar C_2
.
\label{eq:contact2rho}
\end{eqnarray}

\begin{figure}[htb]
\begin{center}
\includegraphics[width=10cm]{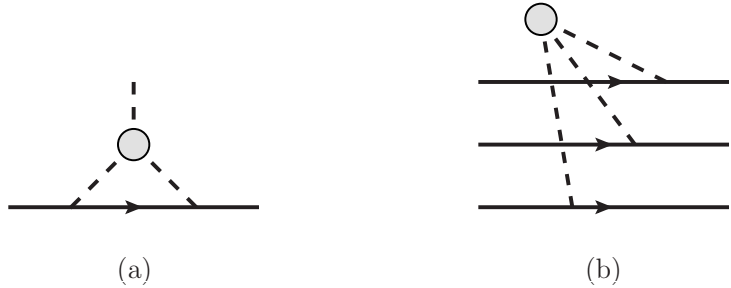}
\caption{\label{fig:three-pion}
Diagrammatic representation of the contribution of the three-pion interaction to the nucleon level CP violating processes.
The solid and dashed lines denote the nucleon and the pion, respectively.
(a) Radiatively generated isovector CP-odd pion-nucleon interaction. (b) CP-odd three-nucleon force.}
\end{center}
\end{figure}

The CP-odd chiral lagrangian (\ref{eq:cpv_chiral_lagrangian}) contains a three-pion interaction which generates the CP-odd three-nucleon force and also additional contribution to the isovector CP-odd pion-nucleon interaction \cite{eft6dim}.
The isovector CP-odd pion-nucleon interaction is radiatively generated by the diagram (a) of Fig. \ref{fig:three-pion}.
This additional contribution, at the leading order, is given by the momentum dependent vertex function
\begin{equation}
\bar g_{3\pi} (k) = 
f_{g_1} (k) \Delta_{3\pi}
,
\label{eq:g_1_3pi}
\end{equation}
with the momentum dependent function \cite{eft6dim}
\begin{equation}
f_{g_1} (k)
=
-\frac{15 g_A^2 m_\pi m_N}{32 \pi f_\pi^2}
\Biggl\{
1 
+ \Biggl[
\frac{1+ 2 \vc{k}^2 / (4m_\pi^2 )}{3 |\vc{k}| / (2 m_\pi)}
{\rm arctan} \Biggl( \frac{|\vc{k}|}{2 m_\pi} \Biggr)
-\frac{1}{3}
\Biggr]
\Biggr\}
,
\label{eq:f_g_1}
\end{equation}
where $k$ is the momentum of the outgoing pion.
The momentum independent part of $f_{g_1}$ (the first term in the curly bracket) gives the leading contribution to the nuclear EDM.
The isovector CP-odd nuclear force is therefore corrected as
\begin{equation}
\bar G^{(1)}_\pi
= 
\frac{g_A m_N }{f_\pi} 
\Biggl[
\frac{
\bar g_1 }{2 f_\pi}
-\frac{15 g_A^2 m_\pi m_N}{32 \pi f_\pi^2} \Delta_{3\pi}
\Biggr]
,
\end{equation}
where we have neglected the momentum dependence.

The three-pion interaction [see Fig. \ref{fig:three-pion} (b)] also generates the CP-odd three-nucleon force \cite{eft6dim}:
\begin{eqnarray}
V_{P\hspace{-.5em}/\, T\hspace{-.5em}/\, }^{3N} (\vc{k}_1 ,\vc{k}_2 ,\vc{k}_3)
&=&
-i \Delta_{3\pi} \frac{m_N g_A^3}{4 f_\pi^3}
\Bigl[
(\vc{\tau}_1 \cdot \vc{\tau}_2 ) \tau^z_3
+(\vc{\tau}_2 \cdot \vc{\tau}_3 ) \tau^z_1
+(\vc{\tau}_3 \cdot \vc{\tau}_1 ) \tau^z_2
\Bigr]
\nonumber\\
&& \hspace{8em} \times
\frac{(\vc{\sigma}_1 \cdot \vc{k}_1 ) (\vc{\sigma}_2 \cdot \vc{k}_2 ) (\vc{\sigma}_3 \cdot \vc{k}_3 )}{\Bigl[ \vc{k}_1^2 + m_\pi^2 \Bigr] \Bigl[ \vc{k}_2^2 + m_\pi^2 \Bigr] \Bigl[ \vc{k}_3^2 + m_\pi^2 \Bigr]}
,
\label{eq:cpv3N}
\end{eqnarray}
where the subscript of the momentum $\vc{k}$, isospin and spin matrices labels the nucleon.
The above CP-odd three-nucleon force depends on the spin and isospin matrices of three nucleons at the same time.
This means that it is active only for configurations where all interacting nucleons are unpaired.
As the nucleons have tendency to be paired in singlet, the effect of the CP-odd three-nucleon force is expected to be small for stable nuclei.

\section{Intrinsic nucleon EDM contribution\label{sec:intrinsicnucleonedm}}

In this Section, we present and analyze the results of the calculations of the intrinsic nucleon EDM contribution to the nuclear EDM (\ref{eq:intrinsicnucleonedm}).
The results of the evaluations of the enhancement factors of the intrinsic nucleon EDM for the deuteron, $^3$He, $^3$H (ab initio using A$v18$ \cite{av18,yamanakanuclearedm}, chiral EFT \cite{bsaisou,bsaisou2}), $^{6}$Li, $^{9}$Be, and $^{13}$C (Cluster model \cite{yamanakanuclearedm,13cedm}) are listed in Table \ref{table:intrinsic_nucleon_edm_nuclearedm}.
Let us now analyze them in detail.

\begin{table}[h]
\tbl{
Enhancement factors of intrinsic nucleon EDM for the deuteron, $^3$He, $^3$H, $^{6}$Li, $^{9}$Be, and $^{13}$C.
The data are quoted from Refs. \citen{yamanakanuclearedm,bsaisou,13cedm}.
For the results of chiral EFT, the error bar is also shown.
}
{
\begin{tabular}{@{}l|cc|cc|cc|cc|@{}} 
\hline
&\multicolumn{2}{|c|}{Ab initio (A$v18$) } &\multicolumn{2}{|c|}{ Chiral EFT } & \multicolumn{2}{|c|}{ Cluster model }&\multicolumn{2}{|c|}{ Simple Shell Model }\\
  &$\langle \sigma_p \rangle$ & $\langle \sigma_n \rangle$    &$\langle \sigma_p \rangle$ & $\langle \sigma_n \rangle$    &$\langle \sigma_p \rangle$ & $\langle \sigma_n \rangle$ &$\langle \sigma_p \rangle$ & $\langle \sigma_n \rangle$     \\ 
\hline
$^{2}$H  & 0.914 & 0.914 & $0.939\pm 0.009$ & $0.939\pm 0.009$ & $-$& $-$&1&1\\
$^{3}$He  & -0.04 & 0.88 & $-0.03\pm 0.01$ & $0.90\pm 0.01$ & $-$& $-$&0&1\\
$^{3}$H & 0.88 & -0.05 & $0.92\pm 0.01$ & $-0.03\pm 0.01$ & $-$& $-$&1&0 \\
\hline
$^{6}$Li  & $-$& $-$ & $-$& $-$& 0.86 & 0.86 &1&1 \\
$^{9}$Be & $-$& $-$ & $-$& $-$& $-$ & 0.75  &0&1 \\
$^{13}$C & $-$& $-$& $-$& $-$& $-$ & $-0.33$  &0&$-\frac{1}{3}$ \\
\hline
\end{tabular}
\label{table:intrinsic_nucleon_edm_nuclearedm}
}
\end{table}

In many-body systems, the EDM of the constituents may be enhanced if their spins are aligned.
It is however known that the nucleons have strong pairing correlations, which tend to minimize the nuclear spin.
Stable even-even nuclei have an angular momentum $0^+$ without exception.
Moreover, there are no stable odd-numbered nuclei with aligned nucleon spin with the same isospin.
We cannot therefore expect enhancements of the nucleon EDM effect from the alignment of the nucleon spin.
Rather, the nuclear spin matrix elements are suppressed by the mixing of angular momentum configurations.
Due to the spin dependent interactions between nucleons, orbital angular momentum configurations with flipped nucleon spin mix together, and their superposition suppresses the nuclear spin matrix elements.
A good example is the intrinsic nucleon EDM contribution to the deuteron EDM, which is smaller than one for both the proton and the neutron (see Table \ref{table:intrinsic_nucleon_edm_nuclearedm}).
This suppression is due to the mixing of $d$-wave, which interferes destructively to the deuteron spin matrix elements \cite{yamanakanuclearedm,bsaisou,bsaisou2}.

Another possible mechanism of enhancement of the EDM of composite systems is the relativistic effects \cite{sandars1,sandars2,flambaum2}.
This process was especially important in the study of the atomic EDM, where the electron EDM is enhanced by the strong Coulomb potential of heavy atoms \cite{ginges}.
This effect is however negligible in nuclear systems, as the nucleons are nonrelativistic, at least for light nuclei which are the main target of this review.
Rather, the nucleon EDM may be partially screened by a mechanism analogue to Schiff's screening \cite{sinoue}, encountered in atoms.
This suppression effect is of the order of percentage for the deuteron, but may be more important for heavier nuclei.
As a general feature, the nuclear spin matrix elements have tendency to become small when the neutron or proton numbers become far from their magic numbers \cite{yoshinaga1,yoshinaga2}.

The enhancement factors of intrinsic nucleon EDM contribution to the nuclear EDM are therefore smaller than one for stable nuclei, and we can affirm that the nuclear EDM generated by the intrinsic nucleon EDM is generally not enhanced.
As we can see in Table \ref{table:intrinsic_nucleon_edm_nuclearedm}, the linear coefficients  of the nucleon EDM contribution to the nuclear EDM ($\langle \sigma_p \rangle$, $\langle \sigma_n \rangle$) are all less than one.
In this sense, we should call them ``spin quenching factors'', rather than ``enhancement factors''.

As a simple way to confirm the reliability of the results, we can estimate the enhancement factors of the nucleon EDM in the simple shell model \cite{ginges,sushkov,sushkov2,sushkov3,fujita}.
In the nuclear shell model, the nucleus is considered as a system made of a core and valence nucleons.
By assuming a spherical nucleus with a valence nucleon $N$, the EDM of odd-numbered nuclei is given by
\begin{equation}
d_A 
= 
d_N \langle \sigma_N \rangle
=
\left\{
\begin{array}{rl}
d_N & (j=\frac{1}{2}+l)\cr
-\frac{j}{j+1} d_N & (j=\frac{1}{2}-l )\cr
\end{array}
\right.
,
\label{eq:simple_shell_model}
\end{equation}
where $j$ and $l$ are the total and valence nucleon orbital angular momenta, respectively.
For even-numbered nuclei with two valence nucleons such as the deuteron or $^6$Li, we just consider that there are two valence nucleons with constructively aligned spins.
By comparing the results of the ab initio and cluster model calculations with the estimation in the simple shell model (see Table \ref{table:intrinsic_nucleon_edm_nuclearedm}), we see that this valence nucleon picture works well in describing the nucleon EDM contribution to the nuclear EDM.
The suppression of the formers against that of the simple shell model results is due to the mixing of angular momentum configurations, as already discussed in this section.

The intrinsic nucleon EDM contribution to the nuclear EDM is important since it can provide effects of isoscalar CP-odd pion-nucleon interaction [see Eq. (\ref{eq:d_1})] on the nuclear EDM, which tends to have stronger isovector dependence ($\bar g_1$).
This is particularly important when we want to probe the effect of the quark EDM or the Weinberg operator.
For detail, see Section \ref{sec:disentangling}.

\section{Ab initio evaluation of the polarization contribution to the EDM of the deuteron, $^3$H and $^3$He\label{sec:abinitio}}

In the {\it ab initio} evaluation of the nuclear EDM, the Schr\"{o}dinger equation of a many-nucleon system is solved using the bare $NN$ potential without any approximations of many-body physics, and the EDMs of the deuteron, $^3$He, and $^3$H nuclei were considered in previous works.
As the phenomenological CP-even realistic nuclear forces, Argonne $v18$ \cite{av18}, Reid 93, Nijmegen II \cite{nijmegen}, and Inside Non-local Outside Yukawa tail (INOY) \cite{inoy} potentials have been used.
For the three-nucleon systems, the effect of the three-body force of Urbana IX \cite{3bodyforce1,3bodyforce2,3bodyforce3,3bodyforce4} was also considered.
Finally, those EDMs were analyzed in the chiral effective field theory (EFT), which is based on the symmetry of QCD.
In this case it is possible to analyze order-by-order the hadron level CP violation, and systematically estimate the error.

The CP-odd interactions are small perturbations, so the nuclear EDM should linearly depend on the them. 
Here we parametrize the EDM in terms of the phenomenological CP-odd $NN$ couplings of Eq. (\ref{eq:CPVhamiltonian}) as
\begin{equation}
d_A 
= 
\sum_{i=0,1} 
\Bigl[
a_{\pi}^{(i)} \bar G_{\pi}^{(i)}
+a_{\eta}^{(i)} \bar G_{\eta}^{(i)}
+a_{\rho}^{(i)} \bar G_{\rho}^{(i)}
+a_{\omega}^{(i)} \bar G_{\omega}^{(i)}
\Bigr]
,
\label{eq:polarizationedmcoef}
\end{equation}
where we take up to isovector terms.
In Tables \ref{table:deuteronedm} and \ref{table:3nucleonedm} are summarized the results of the ab initio calculations of the coefficients $a_X^{(i)}$ $(i=0,1 ; X = \pi , \eta , \rho , \omega)$ of the EDM of the deuteron, $^3$He, and $^3$H nuclei.
The results of the chiral EFT are given from the next-to-next-to-leading order analysis \cite{bsaisou,bsaisou2}.

The most important observation is that the contribution of the pion exchange CP-odd nuclear forces to the deuteron and three-nucleon systems agrees well for the realistic nuclear forces considered.
This is not surprising, because the CP-odd polarization is generated by the long range pion exchange process, and the long range part of the nuclear wave functions is well described by the realistic nuclear forces as well as by the chiral EFT, which respect the pion exchange.
We also see some enhancement of the isoscalar CP-odd nuclear force contribution to the EDM of $^3$He and $^3$H for the results of chiral EFT and INOY potential.
The shift due to the three-nucleon force is less than 10\% (compare A$v18$ only and A$v18$+UIX in Table \ref{table:3nucleonedm}).

The estimation of the theoretical uncertainty is available for the chiral EFT analysis \cite{bsaisou,bsaisou2}.
The error bars of $a_\pi^{(1)}$ for the deuteron and $a_\pi^{(i)}$ ($i=0,1,2$) for the three-nucleon systems are about 10\%.
They were determined from the variance of the results obtained with five different sets of cutoffs (cutoff for the evaluation of the nuclear potential and that needed in the many-body calculation).
For the calculations of the EDM of light nuclei using realistic nuclear potentials, there are no firm way to define the error bar.

The polarization contribution to the deuteron EDM is insensitive to the isoscalar and isotensor CP-odd nuclear forces under the isospin symmetry.
If the isospin breaking effect is considered (e.g. chiral EFT), the isoscalar CP-odd pion-nucleon coupling $\bar g_0$ also contributes to the deuteron polarization, but this has a large theoretical uncertainty \cite{chiral3nucleon}.
This contribution is however smaller than the pion cloud effect of the single nucleon EDM [see Eq. (\ref{eq:d_0})], so we can neglect it.

The isovector CP-odd nuclear force is relevant for the deuteron as well as for the three-nucleon systems.
Here it is important to note that the effect of the three-pion interaction also contributes through the isovector CP-odd pion-nucleon interaction [see Eqs. (\ref{eq:g_1_3pi}) and (\ref{eq:f_g_1})].
Another interesting feature is that the CP-odd three-nucleon interaction generated by the three-pion interaction was found to be subleading in the EDM of $^3$He and $^3$H, compared to the radiative isovector CP-odd nuclear force \cite{bsaisou}.

In regards to the phenomenological CP-odd potential, 
the contributions from the exchange of mesons heavier than the pion ($\eta , \rho , \omega$) are suppressed.
The polarization becomes smaller as the meson mass increases.
If the meson is sufficiently heavy, the contact interaction approximation can be applied.
For those contributions, the results become more dependent on the realistic nuclear force chosen.
The coefficients vary from the first digit, and the theoretical uncertainty is $O(100)\%$ (see Tables \ref{table:deuteronedm} and \ref{table:3nucleonedm})).
This large uncertainty is due to the lack of knowledge of the short range part of the nuclear force.

The effect of heavy meson exchange is integrated out in the chiral EFT analysis, and this effect is renormalized into the contact interactions [see Eq. (\ref{eq:cpv_chiral_lagrangian})].
In Table \ref{table:3nucleonedm}, we have displayed their effects by using the approximations of Eqs. (\ref{eq:contact1eta}) and (\ref{eq:contact2rho}).
The calculation of the effect of contact interactions on the nuclear EDM is affected by a large theoretical uncertainty for the same reason as that of the heavy meson exchange CP-odd nuclear force, and the results are very unstable \cite{bsaisou}.
We should therefore consider their values as the error bar of the EDM calculation.

\begin{table}[h]
\tbl{Comparison of the evaluations of the coefficients $a_X^{(i)}$ $(i=0,1 ; X = \pi , \eta , \rho , \omega)$ of the deuteron EDM.
Coefficients are in unit of $10^{-2}e $ fm.
The data are quoted from Refs. \citen{korkin,chiral3nucleon,yamanakanuclearedm,liu,bsaisou}.
For the results of chiral EFT analysis, the error bar is also shown.
}
{\begin{tabular}{@{}l|cc|cc|cc|cc|cc|@{}}
\hline
$^2$H EDM  &$a_\pi^{(0)}$ & $a_\pi^{(1)}$ 
&$a_\eta^{(0)}$ &$a_\eta^{(1)}$& $a_\rho^{(0)}$ &$a_\rho^{(1)}$ 
&$a_\omega^{(0)}$ &$a_\omega^{(1)}$ \\ 
\hline
Chiral limit  
& $-$ & $1.9 $ 
& $-$ & $-$ &  $-$ &  $-$ 
&$-$ & $-$ \\
A$v18$ 
& $-$ & $1.45 $ 
&  $-$ & $0.157$  & $-$ & $6.25 \times 10^{-2}$ 
&$-$ & $-5.90 \times 10^{-2}$ \\
Reid93 
& $-$ & $1.45 $ 
&  $-$ & $0.168$ & $-$ & $6.83 \times 10^{-2}$ 
&$-$ & $-6.53 \times 10^{-2}$ \\
Nijm II 
& $-$ & $1.47 $ 
&  $-$ & $0.172$ & $-$ & $7.50 \times 10^{-2}$ 
&$-$ & $-7.19 \times 10^{-2}$ \\
Chiral EFT 
& $ ( 4 \pm 5 )  \times 10^{-3}$ & $1.42 \pm  0.13$ 
& $-$ & $-$ &  $-$ &  $-$ 
&$-$ & $-$ \\
\hline
\end{tabular}
\label{table:deuteronedm}}
\end{table}

\begin{table}[h]
\tbl{Comparison of the coefficients $a_X^{(i)}$ $(i=0,1 ; X = \pi , \eta , \rho , \omega)$ of the EDMs of $^3$He and  $^3$H evaluated ab intio with several realistic nuclear forces.
Coefficients are in unit of $10^{-2}e $ fm.
The data are quoted from Refs. \citen{yamanakanuclearedm,song,bsaisou}.
For the results of chiral EFT analysis, the error bar is also shown.
}
{\begin{tabular}{@{}l|cc|cc|cc|cc|cc|@{}}
\hline
$^3$He EDM &$a_\pi^{(0)}$ & $a_\pi^{(1)}$ 
&$a_\eta^{(0)}$ &$a_\eta^{(1)}$& $a_\rho^{(0)}$ &$a_\rho^{(1)}$ 
&$a_\omega^{(0)}$ &$a_\omega^{(1)}$ \\ 
\hline
A$v18$
& $0.59$ & 1.08 
& $-5.77 \times 10^{-2}$ & 0.106& $-3.02 \times 10^{-2}$ & $4.26 \times 10^{-2}$ 
& $2.27 \times 10^{-2}$ & $-5.27 \times 10^{-2}$ \\
A$v18$+UIX 
& $0.55$ & 1.06 
& $-4.78 \times 10^{-2}$ & 0.097& $-2.70 \times 10^{-2}$ & $3.96 \times 10^{-2}$ 
& $1.87 \times 10^{-2}$ & $-5.18 \times 10^{-2}$ \\
Reid93 
& $0.61$ & 1.09 
& $-6.07 \times 10^{-2}$ & 0.115  & $-3.85 \times 10^{-2}$ & $4.87 \times 10^{-2}$ 
& $2.48 \times 10^{-2}$ & $-5.75 \times 10^{-2}$ \\
Nijm II 
& $0.61$ & 1.11 
& $-5.85 \times 10^{-2}$ & 0.123  & $-3.51 \times 10^{-2}$ & $5.42 \times 10^{-2}$ 
& $2.32 \times 10^{-2}$ & $-6.29 \times 10^{-2}$ \\
INOY 
& $1.03$ & 1.09 
& $-0.153$ & 0.155 & $-0.16$ & $8.58 \times 10^{-2}$ 
& $8.34 \times 10^{-2}$ & $-0.103$ \\
Chiral EFT 
& $0.86 \pm 0.10$ & $1.10 \pm 0.15$ 
& $(-5.7 \pm 2.3) \times 10^{-2}$ & $-$ & $(-6.1 \pm 1.5) \times 10^{-2}$ & $-$ 
& $-$ & $-$ \\
\hline
\hline
$^3$H EDM &$a_\pi^{(0)}$ & $a_\pi^{(1)}$ 
&$a_\eta^{(0)}$ &$a_\eta^{(1)}$ & $a_\rho^{(0)}$ &$a_\rho^{(1)}$ 
&$a_\omega^{(0)}$ &$a_\omega^{(1)}$ \\ 
\hline
A$v18$ 
& $-0.59$ & 1.08 
& $5.80 \times 10^{-2}$ & 0.106 & $3.07 \times 10^{-2}$ & $4.27 \times 10^{-2}$ 
& $-2.28 \times 10^{-2}$ & $-5.34 \times 10^{-2}$ \\
A$v18$+UIX 
& $-0.55$ & 1.08 
& $4.78 \times 10^{-2}$ & 0.097& $2.73 \times 10^{-2}$ & $3.96 \times 10^{-2}$ 
& $-1.87 \times 10^{-2}$ & $-5.31 \times 10^{-2}$ \\
Reid93 
& $-0.61$ & 1.11 
& $6.07 \times 10^{-2}$ & 0.116 & $3.93 \times 10^{-2}$ & $4.91 \times 10^{-2}$ 
& $-2.50 \times 10^{-2}$ & $-5.89 \times 10^{-2}$ \\
Nijm II 
& $-0.61$ & 1.13 
& $5.85 \times 10^{-2}$ & 0.124 & $3.54 \times 10^{-2}$ & $5.45 \times 10^{-2}$ 
& $-2.32 \times 10^{-2}$ & $-6.45 \times 10^{-2}$ \\
INOY 
& $-1.03$ & 1.11 
& 0.154 & 0.156 & $0.16$ & $8.62 \times 10^{-2}$ 
& $8.38 \times 10^{-2}$ & $-0.106$ \\
Chiral EFT 
& $-0.84 \pm 0.10$ & $1.08 \pm 0.15$ 
& $(5.6 \pm 2.2) \times 10^{-2}$ & $-$ &$(6.0 \pm 1.5) \times 10^{-2}$ & $-$ 
& $-$ & $-$ \\
\hline
\end{tabular}
\label{table:3nucleonedm}}
\end{table}

\section{Evaluation of the nuclear polarization effect in the cluster model\label{sec:cluster}}

As we have seen in Section \ref{sec:nucleonlevelcpv}, the operator form of the CP-odd nuclear force (\ref{eq:CPVhamiltonian}) is suggesting that the nuclear EDM is enhanced for nuclei with well developed cluster structure.
For that, we have to study the polarization effect of nuclei beyond three-nucleon systems.
In this section, we introduce the cluster model and present the theoretical evaluation of the EDM of $^6$Li, $^9$Be and $^{13}$C.

\subsection{Setup of the cluster model}

The computational cost of calculating the wave function of ${\cal N}$-body systems is known to become exponentially difficult in growing ${\cal N}$.
The evaluation of the nuclear wave functions of light nuclei is already difficult for $^6$Li or $^7$Li.
In nuclear physics, however, the $\alpha$ cluster is very stable, and has a strong correlation even inside nuclei.
For light nuclei, the cluster structure is actually known to be well developed (e.g. the $^6$Li nucleus is given as an $\alpha - p -n $ system).
The idea to treat light nuclei with $\alpha$ clusters has been very successful in describing low lying spectra, and the cluster model has so far extensively been investigated \cite{clusterreview1,clusterreview2,clusterreview3}.
If we model light nuclei in terms of $\alpha$ clusters, the computational cost greatly decreases, and may help us to evaluate their EDM.

To calculate the wave function of light nuclei in the cluster model, we have to determine the interactions involving clusters.
The common concept is to construct an effective $NN$ interaction which respect the model space of the nucleon inside the cluster, and apply the folding.
For the $\alpha -N $ or $\alpha - \alpha$ interactions, the effective interaction is determined from the experimental data of low energy $N-\alpha$ or $\alpha - \alpha$ scattering \cite{schmid,hasegawa,kanada}.
To reproduce the energy spectrum of light nuclei, the phenomenological three- and four-body interactions between $\alpha$ clusters and/or nucleons are also introduced.

Another critically important feature of the cluster model is how to model the effect of Pauli exclusion principle between nucleons of different clusters.
The Pauli exclusion is taken into account via the {\it Orthogonality Condition Model} (OCM), which consists of manually projecting out the forbidden states \cite{ocm1,ocm2,horiuchi1,ocm3,horiuchi2}.
In the few-body cluster model, it is convenient to remove Pauli-forbidden states by including in the hamiltonian the Pauli-blocking operator \cite{Kukulin}
\begin{eqnarray}
V_{\rm Pauli} = \lim_{\lambda \rightarrow  \infty}\ {\lambda}\ \sum_{f}\ {| u_f \rangle}{\langle u_f |}
.
\end{eqnarray} 
For example, the states $f=0S$ and $f=0S,1S,0D$ are removed for the $\alpha-N$ and $\alpha -\alpha$ systems, respectively.
In practice, the forbidden states are approximately given by the energy eigenfunctions of the harmonic oscillator potential.
The coupling $\lambda$ is typically set as $\lambda=10^{4}$~MeV.

By excluding forbidden states, OCM can prevent nuclear clusters from collapsing, and can reproduce the saturation of the nuclear density.
Meanwhile, low lying energy eigenstates with well developed clusters, relevant near thresholds, can also well be described.
The cluster model is therefore very adequate in analyzing the low energy dynamics of light nuclei, comprising shell and cluster correlations.
It has been extensively applied in the study of the structure of multi-$\alpha$ clustered nuclei, such as $^8$Be \cite{funaki8Be}, $^{12}$C \cite{funaki12C}, or $^{16}$O \cite{funaki16O}, and their description was successful.
The model is also well applicable to odd-numbered nuclei with a valence nucleon and several $\alpha$-clusters, such as the $^{13}$C nucleus.
The analysis of the structure of $^{13}$C in the $3 \alpha + N$ cluster model with OCM yields bound states with a shell-like configuration and excited states with an enhanced $^9$Be + $\alpha$ correlation, well reproducing the observed spectrum \cite{yamada13C}.

\subsection{Folding of CP-odd nuclear force in the cluster model}

In calculating the EDM of light nuclei in the cluster model, the CP-odd intercluster force also has to be modeled.
For light nuclei where the $\alpha$ cluster is a relevant degree of freedom, we have to calculate the CP-odd $\alpha-N$ interaction.
Here we present the derivation of the CP-odd $\alpha-N$ interaction by folding the CP-odd $NN$ interaction.
We must note that only the isovector P, CP-odd nuclear force contributes to the nuclear EDM, since the isoscalar and isotensor CP-odd interactions require the spin and isospin flips for both interacting nucleons, which excite the $\alpha$ cluster.
Moreover, all CP-odd nuclear forces vanish for the $\alpha - \alpha$ interaction due to the closed shell.

\begin{figure}[htb]
\begin{center}
\includegraphics[width=12cm]{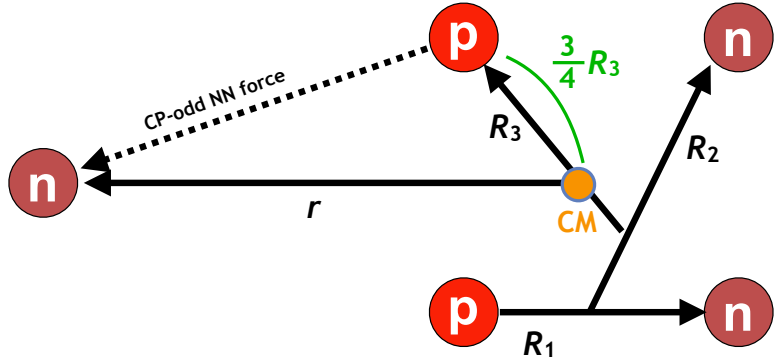}
\caption{\label{fig:cm_shift}
Removal of the spurious center of mass effect in the folding of the CP-odd $\alpha -N$ interaction.
}
\end{center}
\end{figure}

By only considering the direct contribution (no exchange of nucleons between clusters) to the $\alpha-N$ system, the folding of the pion exchange CP-odd nuclear force yields the potential
\begin{equation}
V_{\alpha - N} ( r ) \hat{\vc{r}}
=
\int d^3 \vc{ R}' \, 
V ( m_\pi , |\vc{ r} - \vc{ R}' | )
\rho_\alpha (  R' )
\frac{\vc{ r} - \vc{ R}'}{|\vc{ r} - \vc{ R}' |}
,
\end{equation}
where  $\vc{r}$ is the coordinate of the nucleon with the origin located in the center of mass of the $^4$He nucleus.
The radial function of the CP-odd $NN$ potential is given in Eq. (\ref{eq:barecpvnn}).
To fold the potential we must remove the unphysical effect of the center of mass motion of the $\alpha -N$ system (see Fig. \ref{fig:cm_shift}).
The effective density of the $\alpha$ cluster $\rho_\alpha$, approximated by a gaussian, must therefore be written as
\begin{equation}
\rho_\alpha ( r )
=
4 \, \Biggl( \frac{4}{3} \cdot \frac{\lambda_\alpha}{\pi} \Biggr)^{\frac{3}{2}}
e^{- \frac{4}{3} \lambda_\alpha r^2}
,
\end{equation}
with the oscillator constant $ \lambda_\alpha = ( 1.358 \, {\rm fm})^{-2} = 0.5423$ fm$^{-2}$.
The factor $\frac{4}{3}$ in front of $ \lambda_\alpha$ accounts for the center of mass of the $\alpha$ cluster.

\begin{figure}[htb]
\begin{center}
\includegraphics[width=8cm]{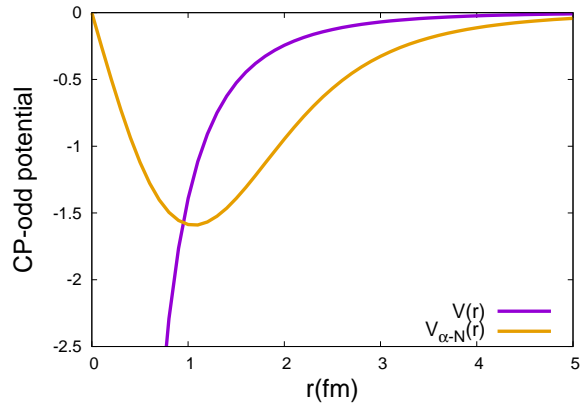}
\caption{\label{fig:folding}
The radial shape of the bare pion exchange CP-odd nuclear force $V(r)$ and its folding potential $V_{\alpha -N}(r)$.
}
\end{center}
\end{figure}

By substituting $ \lambda'_\alpha \equiv \frac{4}{3} \lambda_\alpha $, the folding potential is transformed as
\begin{eqnarray}
V_{\alpha - N} ( r ) \, \hat{\vc{r}}
&=&
-\frac{m_X}{2 \pi m_N}
\Biggl( \frac{\lambda'_\alpha}{\pi} \Biggr)^{\frac{3}{2}}
\int d^3 \vc{R}' \, 
\frac{e^{-m_X |\vc{r} - \vc{R}' | }}{|\vc{r} - \vc{R}' |} \left( 1+ \frac{1}{m_X |\vc{r} - \vc{R}' |} \right)
e^{ -\lambda'_\alpha R'^2 }
\frac{\vc{r} - \vc{R}' }{|\vc{r} - \vc{R}' |}
\nonumber\\
&=&
\frac{m_X}{ m_N}
\frac{\sqrt{\lambda'_\alpha}}{2 \pi^{\frac{3}{2}} r} \hat{\vc{r}} 
\int_0^\infty \hspace{-.7em} d R' \, e^{-m_X R' } \left( 1+ \frac{1}{m_X R' } \right)
\nonumber\\
&&\hspace{3em} \times
\Biggl[ 
e^{ -\lambda'_\alpha (r- R')^2} 
\Biggl( 
\frac{1}{2\lambda'_\alpha r R'}
-1
\Biggr)
- e^{ -\lambda'_\alpha (r+ R')^2} 
\Biggl( 
\frac{1}{2\lambda'_\alpha r R'}
+1
\Biggr)
\Biggr]
.
\label{eq:cpvnn_folding}
\end{eqnarray}
The remaining radial integral cannot be performed analytically.
By numerically integrating the above integral, we obtain the curve of Fig. \ref{fig:folding}.

We see that the CP-odd $\alpha - N$ interaction has a maximal point around 1 fm, and decreases exponentially in growing $r$.
This bump is consistent with the property of the derivative interaction, for which the change of the nuclear density becomes important at the surface, as explained in Section \ref{sec:nucleonlevelcpv}.
The interaction range is also extended due to the nuclear density distribution of the $\alpha$ cluster.
The CP-odd $\alpha - N$ interaction cannot be obtained by naively multiplying the bare CP-odd $NN$ interaction by four due to the above properties.

Here we have to note that in Eq. (\ref{eq:cpvnn_folding}), we have used the bare CP-odd $NN$ interaction.
To be exact, the folding of the bare $NN$ potential is not respecting the model space of the $\alpha -N$ system which has less degrees of freedom, and the use of the effective CP-odd $NN$ interaction is actually required to control the systematics.
It is expected that the long range part of the pion exchange CP-odd nuclear force is little affected by this change of model space, but the short range forces must be modified to a large extent, since they involve energies which may exceed the model space.
It is therefore not possible to treat the CP-odd nuclear forces with heavy meson exchange, or the contact CP-odd interactions in the folding without deriving their effective interactions.
In this review, we only consider the folding of the pion exchange CP-odd nuclear force.

We also point that the isovector CP-odd $\alpha -N$ interaction receives contribution from the three-pion interaction through Eqs. (\ref{eq:g_1_3pi}) and (\ref{eq:f_g_1}).
On the contrary, the CP-odd three-nucleon force (\ref{eq:cpv3N}) is not relevant since the nucleons in the $\alpha$ cluster are in closed shell.
For the study of light nuclei in the cluster model, the latter can therefore be neglected.

\subsection{The EDMs of $^6$Li, $^9$Be, and $^{13}$C\label{sec:edm6li9be13c}}

We now give the result of the evaluation of the EDM of light nuclei in the cluster model.
The nuclear EDMs of $^6$Li, $^9$Be, and $^{13}$C have been calculated so far.
The $^6$Li and $^9$Be were treated as $\alpha - p - n$ and $\alpha - \alpha - n$ three-body systems, respectively \cite{yamanakanuclearedm}.
The $^{13}$C was considered as an $\alpha - \alpha - \alpha - n$ four-body system \cite{13cedm}.
The EDM of those few-body systems were calculated in the Gaussian Expansion Method, which can solve few-body Schr\"{o}dinger equations very accurately \cite{hiyama}.
The results are shown in Table \ref{table:nuclearedm}.
Let us see them in detail one by one.

\begin{table}[ph]
\tbl{Sensitivity of the EDMs of $^6$Li, $^9$Be, and $^{13}$C on nucleon level CP-violating parameters.
The coefficients $a_X^{(i)}$ $(i=0,1,2 ; X = \pi , \eta , \rho , \omega)$ are defined in Eq. (\ref{eq:polarizationedmcoef}).
The results of the ab initio evaluations (A$v18$) of the EDMs of the deuteron, $^3$He, and $^3$H are also shown for comparison.
The coefficients are in unit of $10^{-2}e$ fm.
The data are quoted from Refs. \citen{yamanakanuclearedm,13cedm}.
}
{\begin{tabular}{@{}l|ccc|cc|ccc|cc|cc|@{}} 
\hline
   &$a_\pi^{(0)}$ & $a_\pi^{(1)}$ & $a_\pi^{(2)}$ &$a_\eta^{(0)}$ &$a_\eta^{(1)}$& $a_\rho^{(0)}$ &$a_\rho^{(1)}$ &$a_\rho^{(2)}$  &$a_\omega^{(0)}$ &$a_\omega^{(1)}$ \\ 
\hline
$^{2}$H  & $-$ & $1.45 $ & $-$ &  $-$ & $0.157$ & $-$ & $6.25 \times 10^{-2}$ &  $-$ &$-$ & $-5.90 \times 10^{-2}$ \\
$^{3}$He & $0.59$ & 1.08 & 1.68 & $-5.77 \times 10^{-2}$ & 0.106 & $-3.02 \times 10^{-2}$ & $4.26 \times 10^{-2}$ & $-7.68 \times 10^{-2}$ & $2.27 \times 10^{-2}$ & $-5.27 \times 10^{-2}$ \\
$^{3}$H & $-0.59$ & 1.08 & -1.70 & $5.80 \times 10^{-2}$ & 0.106 & $3.07 \times 10^{-2}$ & $4.27 \times 10^{-2}$ & $7.86 \times 10^{-2}$ & $-2.28 \times 10^{-2}$ & $-5.34 \times 10^{-2}$ \\
$^{6}$Li & $-$ & 2.2 & $-$ & $-$ & 0.17 & $-$ & $7.0 \times 10^{-2}$ & $-$ & $-$ & 0.12 \\
$^{9}$Be & $-$ & $1.4$ & $-$ & $-$ & $-$ & $-$ & $-$ & $- $ & $-$ & $-$ \\
$^{13}$C & $-$ & $-0.20$ & $-$ & $-$ & $-$ & $-$ & $-$ & $- $ & $-$ & $-$ \\
\hline
\end{tabular}
\label{table:nuclearedm}}
\end{table}


The $^6$Li is the lightest stable nuclear system with non-zero total angular momentum which can be found after the three-nucleon systems.
The $^6$Li EDM is well described by the $\alpha + d$ cluster structure, and its EDM inherits the property of that of the deuteron.
The intrinsic nucleon EDM contribution to $^6$Li is very similar to the deuteron.
Concerning the polarization by the CP-odd nuclear force, the $^6$Li is polarized by the deuteron subsystem, and also by the CP-odd $\alpha -N $ interaction.
Those two effects are comparable, and their positive interference enhances the sensitivity of $^6$Li EDM on the isovector CP-odd pion exchange nuclear force.
This case is an example which shows the enhancement of the CP violation by the nuclear cluster structure.

There are additional subdominant effects such as the isoscalar or isotensor CP-odd nuclear forces.
Those effects vanish in the cluster model setup with isospin symmetry.
They may become relevant when the excitation of the $\alpha$ cluster is considered, but this effect requires additional energy of about 20 MeV, which is much larger than the $^6$Li binding energy 3.7 MeV, and therefore negligible.
As for the deuteron, the leading isoscalar effect should be brought by the intrinsic nucleon EDM contribution.

In Ref. \citen{yamanakanuclearedm}, the effect of exchanges of heavier mesons [Eq. (\ref{eq:CPVhamiltonian})] was also considered.
For the $^6$Li, the $\eta$ and $\rho$ exchange CP-odd nuclear forces contribute only to the polarization of the $N-N$ subsystem.
This can be understood by the close values of the isovector coefficients of the $^6$Li nucleus with those of the deuteron.
The $\omega$ exchange, however, also contributes to the polarization of the $\alpha -N$ subsystem.
As it was mentioned previously, a simple folding of heavy meson exchange processes yields results with unknown systematics due to the change of the model space, so we do not consider the $\omega$ exchange.

We also mention the accuracy of this cluster model evaluation.
In the cluster model used, the root mean square distance of the $\alpha -N$ subsystem is about 4 fm.
At this distance, the variation of the CP-odd $\alpha -N$ interaction is mild, and the folding potential should well describe it.
It is expected that the error bar of the isovector CP-odd nuclear force effect does not exceed 20 \%.


The $^9$Be EDM is also analyzed in the same manner.
In the cluster model, the EDM of $^9$Be is polarized only by the CP-odd $\alpha - N$ interaction, and its sensitivity to the isovector CP-odd nuclear force is close to that of the deuteron.
As the distance between an $\alpha$ cluster and the single neutron inside $^9$Be is about 4 fm, the cluster picture is expected to work well, so that the folding of the CP-odd potential describes well the CP-odd polarization.

As the $\alpha$ cluster is considered to be unbreakable, the isoscalar and isotensor CP-odd nuclear forces are not relevant for the polarization contribution to $^9$Be.
For the $^9$Be nucleus, however, the intrinsic neutron EDM also contributes to the nuclear EDM through the spin quenching factor which is close to one  (see Table \ref{table:nuclearedm}).
The leading isoscalar ($\bar g_0$) effect of $^9$Be is therefore generated by the pion cloud process of the neutron EDM.

The effect of heavier meson exchanges cannot be considered in the case of $^9$Be, since the model space of those interactions does not match that of the folding potential.
A similar problem is also encountered for the calculation of the contribution of the CP-odd contact interactions.
In this case, they must be renormalized at the energy scale corresponding to the cutoff of the model space of the $\alpha$ cluster model.


The EDM of $^{13}$C was also evaluated within the cluster model \cite{13cedm}.
As for $^9$Be, the $^{13}$C nucleus is sensitive to the isovector CP-odd pion-nucleon coupling due to the polarization by the CP-odd $\alpha - N$ interaction, and its EDM depends on the isoscalar CP-odd pion-nucleon coupling through the pion cloud effect of the intrinsic neutron EDM.
We do not consider the CP-odd nuclear force due to the exchange of heavier mesons or contact interactions, since the interaction is not renormalized to respect the model space.

The EDM of $^{13}$C is less sensitive on the isovector CP-odd nuclear force than the other lighter nuclei by an order of magnitude.
Its analysis is however interesting in understanding how the CP-odd polarization effect is suppressed.
The $^{13}$C nucleus has a ground state ($\frac{1}{2}^-_1$) which is well understood as a shell-like nucleus.
If we describe it in terms of a system made of a valence nucleon and a $^{12}$C core, the orbital angular momentum $l =1$ and the nucleon spin will be constructively combined due to the spin-orbit force.
The angular momentum of $^{12}$C core therefore has to be $2^+$ to obtain the $\frac{1}{2}^-_1$ state.
The opposite parity excited state of $^{13}$C ($\frac{1}{2}^+_1$) is located 3.1 MeV above the ground state, but this state is a neutron halo state with a $0^+$ $^{12}$C core due to the damp of the spin-orbit force, and the transition between $\frac{1}{2}^-_1$ and $\frac{1}{2}^+_1$ states through the CP-odd nuclear force or the EDM operator is suppressed.

The next candidate of states which can couple to the $\frac{1}{2}^-_1$ state through the parity mixing is the $^{12}$C + $n$ continuum state, which opens at 4.9 MeV above the ground state.
However, we also have here a dominant configuration with 0$^+$ $^{12}$C core, and the transition between them is small.
The state with the best overlap is the continuum state with a neutron and an excited $^{12}$C core (2$^+$, 4.4 MeV above 0$^+$).
Its threshold is 9 MeV above the ground state, and the transition is suppressed energetically.

From the analysis of the EDM of $^{13}$C, we can learn that the existence of low lying opposite party states with the same angular momentum is not a sufficient condition to obtain large CP-odd nuclear polarization.
The bad transition between low lying opposite parity states may also be relevant for $^{15}$N, which have a $\frac{1}{2}^-$ ground state and a $\frac{1}{2}^+$ state at 5.3 MeV.
This analysis is also suggesting that the EDM of nuclei with a shell-like structure is less advantageous than clustered nuclei in enhancing the CP-odd effect.

Let us add a brief comment on the effect of the CP-odd contact interaction of chiral EFT.
The isospin blind contact interaction [term with $\bar C_1$ of Eq. (\ref{eq:CPVEFTpot})] in principle does contribute to the CP-odd $\alpha -N$ interaction.
The contribution of $\bar C_1$ to the EDM of $^6$Li is forbidden by isospin selection rule.
The $^9$Be and $^{13}$C nuclei can probe the effect of $\bar C_1$.
To evaluate it correctly, we have to derive the effective interaction by changing the model space, but it is not currently available.
The isospin dependent isoscalar CP-odd contact interaction [term with $\bar C_2$ of Eq. (\ref{eq:CPVEFTpot})] cannot contribute to the CP-odd $\alpha -N$ folding potential, since it has to open the isospin shell of the $\alpha$ cluster.
Therefore, $\bar C_2$ cannot be probed with the EDMs of $^6$Li, $^9$Be, $^{13}$C.

\section{$\theta$-term contribution\label{sec:theta}}

It is known that the strong interaction involves a CP violating interaction, the $\theta$-term, defined by
\begin{equation}
{\cal L}_{\theta} 
=
\bar \theta \, \frac{\alpha_s}{8 \pi} \epsilon^{\mu \nu \rho \sigma} G_{\mu \nu}^a G_{\rho \sigma}^a 
,
\label{eq:theta-term}
\end{equation}
where $\alpha_s$ is the strong coupling and $G_{\mu \nu}^a$ the gluon field strength.
This term is not forbidden in SM, so it is of importance to quantify its effect on the hadronic CP violating observables.

The $\theta$-term contribution to the nucleon EDM was evaluated in many phenomenological approaches \cite{crewther,pich,borasoy,narison,pospelovtheta1,pospelovtheta2,pospelovreview,nedmholography}.
The chiral EFT analysis of the nucleon EDM within the $\theta$-term involves several low energy constants which have to be fitted from phenomenology or lattice QCD data [see Eqs. (\ref{eq:d_0}) and (\ref{eq:d_1})].
Using lattice QCD data with pion mass out of its physical one \cite{shintani2}, the nucleon EDMs calculated in chiral EFT are \cite{ottnad,mereghetti2,mereghetti1,guo1}
\begin{eqnarray}
d_n
&= &
-(2.7 \pm 1.2) \times 10^{-16} \bar \theta \,
e\, {\rm cm}
,
\\
d_p 
&=&
(2.1 \pm 1.2) \times 10^{-16} \bar \theta \,
e\, {\rm cm}
.
\end{eqnarray}
Continuous efforts to calculate the nucleon EDM in lattice QCD are also on-going, although simulations at the physical pion mass are still difficult \cite{nedmlattice1,shintani1,nedmlattice3,shintani2,nedmlattice5,nedmlattice6,nedmetm,shintani3}.
Currently, results of the calculation of the nucleon EDM at the pion mass $m_\pi = 170$ MeV are giving \cite{shintani3}
\begin{eqnarray}
d_n
&= &
-(0.93 \pm 0.43) \times 10^{-14} \bar \theta \,
e\, {\rm cm}
,
\\
d_p 
&=&
(1.01 \pm 0.90) \times 10^{-14} \bar \theta \,
e\, {\rm cm}
.
\end{eqnarray}
The direct calculation of the $\theta$-term contribution to the nucleon EDM at the physical pion mass $m_\pi = 135$ MeV is one of the important goal in the study of hadronic CP violation.

The most recent experimental data of the neutron EDM is giving an upper limit \cite{baker}
\begin{equation}
d_n < 2.9 \times 10^{-26}e\, {\rm cm}
.
\label{eq:neutron_edm_exp_data}
\end{equation}
The constraint on the $\theta$-parameter is then
\begin{equation}
\bar \theta < 10^{-10}
.
\label{eq:thetalimit}
\end{equation}
As this term is not suppressed or forbidden by some symmetries or mechanisms in SM, the natural size of the coupling should be of order one, like the CP-even QCD lagrangian.
The constraint from the experimental data is suggesting a much smaller $\bar \theta$ than its naturally expected size.
This problem is known as the {\it Strong CP Problem}.

As a natural resolution of this problem, a mechanism making the $\theta$-term irrelevant was proposed by Peccei and Quinn \cite{peccei}.
This consists of introducing a new scalar field, the axion, which has the following lagrangian
\begin{equation}
{\cal L}_a = \frac{1}{2} \partial_\mu a \partial^\mu a + \frac{a(x)}{f_a} \frac{\alpha_s}{8\pi} \epsilon^{\mu \nu \rho \sigma} G_{\mu \nu}^a G_{\rho \sigma}^a
,
\label{eq:axionlagrangian}
\end{equation}
with $f_a$ the decay constant of some symmetry which is broken at some energy scale much higher than the electroweak one.
This field develops an expectation value at $\langle a \rangle / f_a+  \bar \theta = 0$ due to the nonzero topological susceptibility of QCD, and the effect of the $\theta$-term is dynamically canceled.
This attractive mechanism was often considered in the study of hadronic CP violation \cite{pospelovreview}.

The axion mechanism cancels the effect of the bare $\theta$-term, but it may also induce dynamical one in the presence of additional CP-odd operators, through the correlation with the topological charge \cite{bigi1,bigi2}.
As an example, the induced $\theta$-term due to the quark chromo-EDM is
\begin{equation}
\bar \theta_{\rm ind} 
= 
\frac{m_0^2}{2} \sum_{q} \frac{d^c_q}{m_q} \ .
\label{eq:inducedthetacedm}
\end{equation}
where $m_0^2 \equiv -\frac{\langle 0 | g_s \bar q \sigma_{\mu \nu} t_a G_a^{\mu \nu} q | 0 \rangle}{\langle 0 | \bar q q | 0 \rangle} =0.8\, {\rm GeV}^2$ \cite{belayev1,belayev2}.
This induced $\theta$-term has comparable contribution as other hadronic CP-odd effective interactions, and cannot be neglected.
This means that the evaluation of the $\theta$-term contribution to hadronic CP violation is important even if the axion mechanism is active.

If the axion mechanism is inactive, Eq. (\ref{eq:thetalimit}) is a phenomenological constraint that corresponds to the theoretical uncertainty of the SM contribution.
In the analysis of the new physics contribution to the nuclear EDM, the effect of the $\theta$-term must therefore be investigated within the allowed region \cite{bsaisou,bsaisou2,devriesdeuteron,bsaisoudeuterontheta,mereghettitheta}.
As we have seen previously, the contribution of the CP-odd nuclear force is important in the study of the EDM of light nuclei.
The $\theta$-term contributes dominantly to the isoscalar CP-odd pion-nucleon interaction, and this contribution can be accurately evaluated in the chiral approach.
The most recent prediction based on flavor $SU(3)$ interactions with $SU(3)$ splittings is giving \cite{devriessplitting}
\begin{equation}
\frac{\bar g_0 (\bar \theta)}{2 f_\pi} 
=
(15.5 \pm 2.5) \times 10^{-3} \bar \theta
.
\end{equation}
For light nuclei, the effect of isovector CP-odd pion-nucleon interaction is often more important.
The $\theta$-term, although being isoscalar, also generates sizable isovector coupling $\bar g_1$.
The next-to-leading order analysis in chiral $SU(3)$ EFT is giving \cite{devriessplitting}
\begin{equation}
\frac{\bar g_0 (\bar \theta)}{\bar g_1 (\bar \theta)}
\sim
5
.
\end{equation}
The other hadron level CP violation such as the three-pion interaction or the contact CP-odd nucleon-nucleon interaction, generated by the $\theta$-term, has negligible effects.
In Table \ref{table:smnuclearedm}, we show the uncertainty of the EDM of several light nuclei when the $\theta$-term respects the experimental upper limit by the neutron EDM (\ref{eq:neutron_edm_exp_data}).
We see that the nuclear EDM is typically of the same order as that of the neutron EDM.

\begin{table}[htbp]
\tbl{Nuclear EDM in SM for several nuclei.
The uncertainty of the nuclear EDMs due to the $\theta$-term respecting the constraint from the neutron EDM experimental data \cite{baker} and the CKM contribution are shown.
The uncertainty of the nuclear EDM due to the $\theta$-term is given in absolute values.
}
{\begin{tabular}{@{}l|r|r|c|@{}} 
\hline
   & Uncertainty due to $\theta$-term & CKM prediction & Experimental prospect  \\  
\hline
\ $n$   & $ 2.9 \times 10^{-26}e$ cm  & $(1-6) \times 10^{-32}e$ cm & $\sim 10^{-28}e$ cm \\ 
\ $p$   & $ 2 \times 10^{-26}e$ cm & $(1-6) \times 10^{-32}e$ cm &$\sim 10^{-29}e$ cm \\ 
$^{2}$H & $ 1 \times 10^{-26}e$ cm & $ 3 \times 10^{-31}e$ cm &$\sim 10^{-29}e$ cm \\ 
$^{3}$He & $ 5 \times 10^{-26}e$ cm & $ 3 \times 10^{-31}e$ cm &$\sim 10^{-28}e$ cm \\ 
$^{3}$H  & $ 3 \times 10^{-26}e$ cm & $ 0.7 \times 10^{-31}e$ cm &$\sim 10^{-28}e$ cm \\ 
$^{6}$Li & $ 1 \times 10^{-26}e$ cm & $ 4 \times 10^{-31}e$ cm &$-$ \\ 
$^{9}$Be & $ 1 \times 10^{-26}e$ cm & $ 2 \times 10^{-31}e$ cm &$-$ \\ 
$^{13}$C & $ 0.3 \times 10^{-26}e$ cm & $ -0.3 \times 10^{-31}e$ cm &$-$ \\ 
\hline
\end{tabular}
\label{table:smnuclearedm}}
\end{table}

\section{Nuclear EDM from the CP violation of the Cabibbo-Kobayashi-Maskawa matrix\label{sec:ckmedm}}

The EDM is attracting interest in part due to the small CKM contribution, which allows us to get rid of it in the background analysis.
Here we are considering the case where the $\theta$-term is not relevant, for instance when it is removed by the axion mechanism.
In the case of the nuclear EDM, simple estimation of CKM effect gives $d_A \sim O(\frac{\alpha_s}{4\pi} G_F^2 J \Lambda_{\rm QCD}^3 ) \sim 10^{-32}e$ cm, with the Jarlskog invariant $J = (3.06^{+0.21}_{-0.20} ) \times 10^{-5}$ \cite{pdg,jarlskog}, while earlier calculations are predicting nuclear EDMs in the range $O(10^{-30} - 10^{-33}) e$ cm \cite{avishai1,avishai2,avishai3,donoghue1,hesmedm}.
At the quark level, there are several short distance processes which induce the CP violation due to the CKM phase, such as the $\theta$-term \cite{ellisthetasm,khriplovichtheta,gerard}, the quark chromo-EDM \cite{czarnecki}, the Weinberg operator \cite{smweinbergop}, etc.
Those effects are however known to be small.
Here we discuss the nuclear EDM induced by the CKM CP phase, which may be enhanced by the nuclear many-body effect.

\begin{figure}[htb]
\begin{center}
\includegraphics[width=10cm]{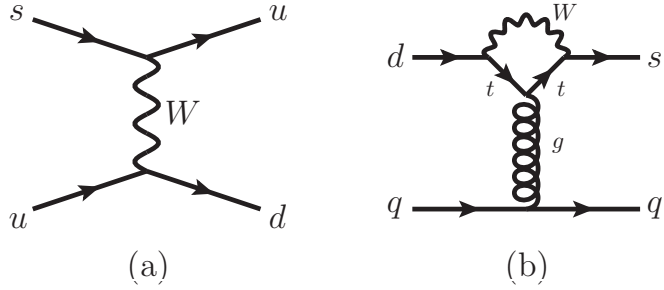}
\caption{\label{fig:deltas1}
Leading CKM contributions to the quark level CP violation.
(a) Tree level diagram. (b) Penguin diagram.
}
\end{center}
\end{figure}

Here we present current results of the calculations of the nuclear EDM induced by the CKM phase through the long distance process \cite{yamanakasmedm}.
The leading CKM CP violation is generated by the tree level $|\Delta S|=1$ four-quark interaction (with the product of CKM matrix elements $V_{us}^* V_{ud}$) and by the penguin diagram (with $V_{ts}^* V_{td}$) \cite{sushkov}.
The combination of those elements forms the Jarlskog invariant (see Fig. \ref{fig:deltas1}).
Those quark level processes are each matched with different hadron level CP violating interactions.
The tree level $|\Delta S|=1$ interaction is matched with the hyperon-nucleon interaction, often using the quark model.
The penguin diagram contribution is matched with the $|\Delta S|=1$ P-odd meson-baryon interaction.
Here the factorization model with the vacuum saturation approximation is often used.
It is important to note that the penguin contribution is enhanced by an order of magnitude through the renormalization group equation when its Wilson coefficient is evolved from the electroweak scale to the hadronic scale \cite{buras,buras2,yamanakasmedm}.

\begin{figure}[htb]
\begin{center}
\includegraphics[width=10cm]{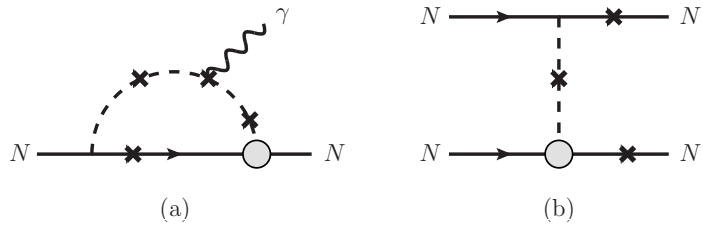}
\caption{\label{fig:smcpv}
Leading CKM contributions to the nucleon level CP violation.
The inner solid lines are either a nucleon or a hyperon ($\Lambda$ or $\Sigma$).
The dashed lines denote mesons ($\pi$, $K$ or $\eta$). 
(a) Pion loop contribution to the nucleon EDM. (b) Pion exchange CP-odd nuclear force.
The grey blob denotes the $|\Delta S|=1$ interaction induced by the penguin diagram.
The crosses denote the possible insertions of $|\Delta S|=1$ interactions generated by the tree level diagram.
}
\end{center}
\end{figure}

After obtaining the hyperon-nucleon transition and the $|\Delta S| =1$ P-odd meson-baryon interactions, we can combine them to calculate the nucleon level CP violation.
The CP phase of the CKM matrix becomes relevant at the hadron level.
The leading nucleon level CP violation is given by the nucleon EDM [see Fig. \ref{fig:smcpv} (a)] and the CP-odd nuclear force [see Fig. \ref{fig:smcpv} (b)].
Here we must note that hyperons and kaons have to be explicitly treated as dynamical degrees of freedom to evaluate the long distance effect (we have to enlarge the cutoff).
In Table \ref{table:smnuclearedm}, we show the estimated values of the EDMs of light nuclei generated by the CKM CP phase.
In these calculations, the polarization effect due to the exchange of $\pi$, $K$, and $\eta$ mesons was considered \cite{yamanakasmedm}.
For the deuteron, an additional dynamical effect due to the $NN - \Lambda N - \Sigma N$ channel coupling was also considered, and a deviation of 10\% was found \cite{smdeuteronedm}.

The EDM of light nuclei induced by the CKM CP violation are in the general case below the prospective sensitivity of the prepared measurement using storage rings [$d_A \sim O(10^{-29})e$ cm].
In this sense, we can safely neglect CKM backgrounds in the search for new physics BSM using nuclear EDMs.
The evaluation of the nuclear EDM involves however large systematics which are difficult to control.
The largest theoretical uncertainty comes from the QCD calculation of the weak $|\Delta S|=1$ meson-baryon couplings.
From the large $N_c$ analysis, this error is estimated to be $O(100\%)$.
The accuracy of the contribution of the $|\Delta S|=1$ four-quark interactions to the $|\Delta S|=1$ interbaryon potential is expected to be improved through the phenomenological and EFT analyses of the nonleptonic weak decays of hyperons \cite{donoghue2,donoghue3,he1,hesmedm,hiyamahyperondecay} and hypernuclei \cite{inoue1,inoue2,sasaki1,sasaki2,perez1,perez2,perez3}.

The nucleon EDM contribution to the nuclear EDM is also an important source of uncertainty.
The long distance effect of the nucleon EDM from the CKM CP phase was estimated to be $O(10^{-32})e$ cm in many previous works \cite{ellis,khriplovichnedm,mckellar1,seng}, an order of magnitude smaller than the EDM of light nuclei (see Table \ref{table:smnuclearedm}).
The contribution of the nucleon EDM is not enhanced inside the nucleus, due to the spin quenching factor smaller than one (see Section \ref{sec:intrinsicnucleonedm}).
The uncertainty is however enlarged due to unknown relative sign with the nuclear EDM \cite{seng}.

We should note that the tree level CKM contribution with higher dimension operators was also evaluated, yielding a nucleon EDM of $d_n \sim O(10^{-31})e$ cm \cite{mannel}.
There the baryon matrix elements were estimated using the naive dimensional analysis, with a suppression factor of $\frac{1}{3}$ due to the strange quark.
However, from recent lattice QCD analyses of nucleon matrix elements such as the nucleon strange content or the axial charge, it is known that the strange quark effect is smaller by one or two orders of magnitude \cite{qcdsf1,engelhardt,jlqcd2,jlqcd3,junnarkar,gong,etm1,etm2,jlqcd4,chiqcdsigmaterm,rqcdsigmaterm,bmwsigmaterm,etmsigmaterm}.
This result for the nucleon EDM should therefore be recognized as the upper limit of the theoretical uncertainty.



\section{Prospects for the search of new physics beyond standard model\label{sec:prospects}}

\subsection{Prospects for several candidate models}

\begin{figure}[htb]
\begin{center}
\includegraphics[width=12cm]{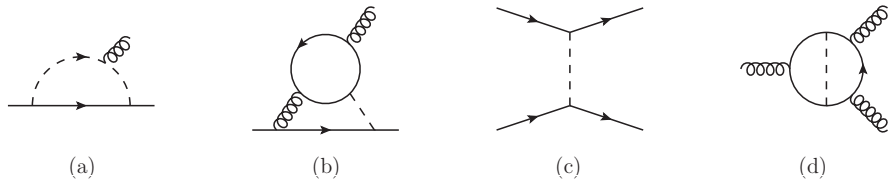}
\caption{\label{fig:tev_cpv}
Diagrammatic representation of several known important elementary level CP violating processes contributing to the nuclear EDM.
The dashed lines denote the boson of new physics BSM.
(a) One-loop level quark chromo-EDM, (b) Barr-Zee type two-loop level diagram, (c) CP-odd four-quark interaction, (d) Weinberg operator.
}
\end{center}
\end{figure}


Let us now see the prospects for the discovery of new physics BSM.
After the integration of new particles, the CP violation BSM generates several dimension-six operators.
The leading CP violatiing lagrangian is given by
\begin{eqnarray}
{\cal L}_{\rm CPV}
&=&
- \sum_q d_q \frac{i}{2} \bar q \sigma_{\mu \nu} F^{\mu \nu} \gamma_5 q
- \sum_q d^c_q \frac{i}{2} \bar q \sigma_{\mu \nu} G^{\mu \nu}_a \gamma_5 t_a q
\nonumber\\
&&
+\frac{1}{6} w f^{abc} \epsilon^{\alpha \beta \gamma \delta} G^a_{\mu \alpha } G_{\beta \gamma}^b G_{\delta}^{\mu,c}
+\sum_{q , q' } \sum_{k} C_{q q' ,k} D_{\alpha \beta \gamma \delta , k}
\bar q_\alpha i\gamma_5 \Gamma_k q_\beta \, \bar q'_\gamma \Gamma_k q'_\delta
,
\nonumber\\
\label{eq:tevcpv}
\end{eqnarray}
where $\Gamma = \hat{1}$ or $\sigma_{\mu \nu}$ and $D_{\alpha \beta \gamma \delta} = \delta_{\alpha \beta} \delta_{\gamma \delta}$ or $\delta_{\alpha \delta} \delta_{\gamma \beta}$.

We first discuss the case of supersymmetric models \cite{pospelovreview,mssmreloaded}.
In supersymmetric models, the fermion EDM [the first term of Eq. (\ref{eq:tevcpv})] and the chromo-EDM [the second term of Eq. (\ref{eq:tevcpv})] receive contribution from the one-loop level \cite{ellismssmedm} [see Fig. \ref{fig:tev_cpv} (a)].
By assuming that the quark chromo-EDM contributes to the leading order, the sensitivity of the EDM of light nuclei of $O(10^{-29})e$ cm can probe $\theta_\mu$ and $\theta_A$, the CP phase of the $\mu $ term and that of the trilinear supersymmetry breaking coupling, respectively, at the level of $O(10^{-2})$ in supersymmetric models with the supersymmetry breaking scale $M_{\rm SUSY} \sim$ TeV \cite{pospelovreview}.
Here we have assumed the axion mechanism \cite{peccei} is active and $\tan \beta = O(1)$.
Even if the phases $\theta_\mu$ and $\theta_A$ are flavor independent, the current quark mass splitting generates both isoscalar and isovector contributions to the chromo-EDM, and consequently the isoscalar and isovector CP-odd pion-nucleon interactions $\bar g_0$ and $\bar g_1$.
Therefore, light nuclei which have high isovector sensitivity can probe the supersymmetric CP phases.
The sensitivity to $\theta_\mu$ is increased with large $\tan \beta$ \cite{pospelovreview,demir}.
This prospective sensitivity may unveil the high scale supersymmetry breaking scenario, which is beyond the sensitivity of Large Hadron Collider (LHC) experiment.


We now model the effect of CP violation BSM by the CP-odd four-quark interaction [the fourth term of Eq. (\ref{eq:tevcpv}), see Fig. \ref{fig:tev_cpv} (c)] with the exchange of a new boson with mass $M_{NP}$ and with $O(1)$ CP phase for the interactions among bosons and light quarks.
A typical isovector CP-odd pion-nucleon interaction with the coupling $\bar g_1 \sim g^2_{NP} \frac{\Lambda_{\rm QCD}^2}{M_{NP}^2} $ is therefore generated in the factorization approach with vacuum saturation approximation.
In this case $\bar g_1$ is not suppressed by the light quark mass.
This contribution is important when the boson is exchanged between two light quarks.
Here the QCD scale parameter is $\Lambda_{\rm QCD} \sim 200$ MeV, and $g_{NP}$ is the coupling between quarks and new particles.
Another important possibility is the generation of the three-pion interaction, which also contributes to $\bar g_1$ and can be probed at the same level.
If the EDM of light nuclei can be measured at the level of $O(10^{-29})e$ cm, we can probe the new physics of the energy scale $M_{NP} \sim$ PeV, with a typical coupling $g_{NP} = O(0.1)$.
This estimation works for models which generate CP-odd 4-quark interactions, such as the Left-right symmetric model \cite{xu,dekens2,nemev} or the sfermion exchange processes in R-parity violating supersymmetry \cite{yamanakabook,faessler}.
We have to note that $g_{NP}^2$ may be suppressed by factors of light quark mass ($m_{u,d}/\Lambda_{\rm QCD} \sim 10^{-2}$) when the model considered involves a Yukawa like coupling.
In this case, the constraint on the scale of new physics will be attenuated, inversely proportional to $g_{NP}^2$.


Another important class of models is that inducing the Barr-Zee type diagrams [see Fig. \ref{fig:tev_cpv} (b)].
This contribution is generated by the exchange of some scalar boson between a light quark and a heavy particle, and it is complementary to the CP-odd four-quark interaction.
The combined analysis of the CP-odd four-quark interaction and the Barr-Zee type process is therefore important in the analysis of flavor physics.
Typical models contributing to the Barr-Zee type diagram are the Higgs doublet models \cite{2higgs,chang2hdm1,abe}, generic supersymmetric models \cite{chang,pilaftsis,chang2,demir,yamanakarainbow1}, and those with R-parity violation \cite{rpvyamanaka1,rpvyamanaka2,rpvyamanaka3}. 
The Barr-Zee type diagram with the gluon exchange gives the most important effect to the nuclear EDM.
Using the simple formula $d^c_q \sim \frac{m_Q \alpha_s Y_q Y_Q}{16 \pi^3 m_{NP}^2} \ln \frac{m_Q^2}{m_{NP}^2}$, the sensitivity of the nuclear EDM of $O(10^{-29})e$ cm can probe the energy scale of the new physics of $M_{NP} \sim \sqrt{Y_Q} $ TeV.
Here $m_Q$ is the mass of the inner loop particle $Q$, and $Y_Q$ is the coupling between the exchanged scalar boson and $Q$.
We have assumed that the light quark Yukawa coupling is $Y_q \sim 10^{-5}$.
The Barr-Zee type diagram is suppressed by the Yukawa coupling $Y_q$ which reflects the light quark mass.


In some models of new physics BSM, only the quark EDM is relevant.
This is the case of split supersymmetry \cite{arkani-hamed,chang3,giudice,dhuria}, R-parity violation involving light quarks and heavy leptons \cite{yamanakabook,rpvyamanaka3}, or models generating Barr-Zee type diagrams with charged Higgs, $W$ or $Z$ bosons \cite{bowser-chao}.
Here we do not have to consider other CP-odd operators than the quark EDM, since it does not mix with the others in the leading order renormalization group evolution \cite{degrassi,yang,dekens1,tensorrenormalization1,tensorrenormalization2}.
For those models, the only relevant hadron level CP violating process is the nucleon EDM, for which the quark EDM dominantly contributes.
The quark EDM effect to the nucleon EDM is suppressed by the nucleon tensor charge due to the dynamical effect of QCD \cite{yamanakasde1,pitschmann,bacchetta,courtoy,radici,kang,gutsche,yez}.
Recent lattice QCD data are giving $d_n \approx 0.8 d_d - 0.2 d_u$ \cite{green,rbcukqcdisovectortensor,rqcdisovector,chiqcdisovector,pndmeisovector,etmisovector,pndmetensor1,pndmetensor2,jlqcd4}.
We must also note that the Wilson coefficient of the quark EDM operator is suppressed in the change of scale by the renormalization group evolution \cite{degrassi,yang,dekens1,tensorrenormalization1,tensorrenormalization2} (typically, about 80\% when we run from $\mu =1$ TeV to $\mu = 1 $ GeV).
Moreover, the intrinsic nucleon EDM contribution to the nuclear EDM is smaller than one, due to the spin quenching of the nuclear spin matrix element (see Section \ref{sec:intrinsicnucleonedm}).
This last property is indicating that the nuclear EDM is not an optimal probe of models BSM which only induce the quark EDM.


We finally have models contributing to the Weinberg operator [the third term of Eq. (\ref{eq:tevcpv}), see Fig. \ref{fig:tev_cpv} (d)].
This effect was first discussed for the case of the Higgs doublet model \cite{weinbergop}.
A recent important issue is the CP violation of models with vectorlike quarks and bosons \cite{choi}.
From the dimensional analysis, the Weinberg operator contributes to the nucleon EDM and to the isoscalar contact CP-odd nucleon-nucleon interaction without suppression by the light quark mass, whereas the CP-odd pion-nucleon interactions suffer from it.
The analysis of the nucleon EDM within QCD sum rules is estimating the Weinberg operator contribution as $d_N \sim w \times 20\, e $ MeV \cite{pospelovweinbergop}, where $w$ is the Weinberg operator coupling.
We must note that the running of the Weinberg operator reduces $w$ by an order of magnitude in the evolution of the scale from 1 TeV to 1 GeV, while the quark EDM and the chromo-EDM with Wilson coefficients of the same order are generated due to the operator mixing \cite{degrassi,yang,dekens1}.
If we simply assume $w \sim \frac{g_{NP}^2}{(4 \pi m_{NP})^2}$ is the typical coupling, the sensitivity of the nuclear EDM of $O(10^{-29})e$ cm can probe a model candidate of new physics with $m_{NP} \sim O(100)$ GeV, with a typical coupling $g_{NP} = O(0.1)$.
We must note that the Weinberg operator dominantly contributes to the nucleon EDM, so the EDM of light nuclei, being sensitive to the isovector CP-odd pion-nucleon interaction, is not the most appropriate in probing it.

\subsection{The role of the nuclear many-body effect in disentangling new physics beyond standard model\label{sec:disentangling}}

By summarizing the sensitivity of the nuclear EDM on CP violating hadron level parameters, we see that the contribution from the nucleon EDM becomes smaller than that of the isovector CP-odd pion-nucleon coupling as the nucleon number grows.
As we have seen, the intrinsic nucleon EDM effect cannot be enhanced due to the pairing of nucleons, and it rather has tendency to decrease due to the mixing of angular momentum configuration.
The CP-odd processes which can only be probed by the nucleon EDM, such as the quark EDM, therefore become subdominant.


The isoscalar CP-odd nuclear force also has less effects on the nuclear EDM than the isovector one for light and stable nuclei.
For nuclei with the same proton and neutron numbers the contribution from the isoscalar CP-odd nuclear force becomes suppressed in the nuclear polarization \cite{chiral3nucleon}.
For asymmetric p-shell nuclei, there is also a suppression due to folding of the CP-odd $NN$ interaction.
The complete cancellation of the isoscalar CP-odd nuclear force by the folding may be violated by the dynamical configuration mixing of nucleons, but this effect should be small, since the cluster model describes well the energy spectrum of light nuclei.
The largest part of the sensitivity to $\bar g_0$ is often brought by the pion cloud effect of the valence nucleon, which is not large.
The isoscalar component of the new physics contributing to the $\theta$-term (see Section \ref{sec:theta}) and to the quark chromo-EDM \cite{pospelovcpvgpinn,eft6dim} suffers from this suppression.
For heavy nuclei, the isoscalar CP-odd interaction has sizable effect, because the number of neutrons is well larger than the proton one, and also because the configuration mixing of valence nucleons is important \cite{yoshinaga2}.


The large sensitivity of the nuclear EDM on the isovector pion-nucleon coupling makes light nuclei to be good probes of new physics contributing to the isovector processes.
Up to mass dimension-6 operators, those are the isovector component of the quark chromo-EDM and the isovector CP-odd four-quark interaction relevant in Left-right symmetric models \cite{devries1,dekens1}.
It is useful to note that those two CP violating interactions do not mix with each other in the renormalization group evolution \cite{dekens1}.
This property is very remarkable because those contributions can be analyzed on a priority basis, without caring destructive interference with other CP-odd sources.
This also means that the accurate hadron level calculation of $\bar g_1$ from either the quark chromo-EDM or the CP-odd four-quark interaction is not required as long as we are intending to discover the existence of CP violation BSM without going into investigations of the coupling constants beyond the second digit.
This way of analyzing should work in a natural scenario where CP violating couplings are not fine-tuned.


Another very important role of measuring the EDMs of several nuclei is to fix the CP violating hadron level CP violating coupling constants using the linear independence of the EDMs on them.
This analysis works for scenario with fine-tuned coupling constants.
To fix unknown variables, we need the same number of equations, so to determine the seven unknown couplings of the leading CP violating chiral lagrangian (\ref{eq:cpv_chiral_lagrangian}), we need at least seven measurements of nucleon and nuclear EDMs.
In this procedure, the individual data of the proton and neutron EDMs as well as those of atomic EDMs \cite{ginges,dobaczewski}, which are more sensitive on $\bar g_0$, can also be used.
The experimental measurement of T-odd neutron-nucleus scattering \cite{gudkovreview,bunakov,bowman} may also provide very useful informations on $\bar g_0$, since this process does not depend on the intrinsic nucleon EDM.
Several investigations trying to constrain the multidimensional parameter space of the new physics with the experimental data of the EDMs of various systems (neutron, atoms, and molecules) exist in the literature \cite{yamanakabook,ibrahim,ramseyli,ellisgeometric,ellisgeometric2,rpvlinearprogramming}.
Adding the experimental results of the nuclear EDM measurement will additionally constrain the degrees of freedom of the parameter space of new physics.
The dependence of the atomic and nuclear systems on the hadronic CP violation is complementary, since interesting atoms are generally heavy, and have sizable isoscalar sensitivity, whereas light nuclei are good in probing isovector CP-odd interactions.
We have to note that this multidimensional analysis requires the hadronic CP violation to be accurately quantified.
In the study of atomic systems, the nuclear level evaluations become also problematic, in addition to the hadron level difficulty already relevant in the study of the nuclear EDM.
To fix the elementary level parameters within reasonable accuracy, further improvements of the QCD and nuclear level calculations of the CP violating effects are required.


Finally, the determination of the Weinberg operator has a critical difficulty.
We have seen previously seen that the Weinberg operator mixes with the quark EDM and chromo-EDM when the operator run from the TeV scale to the hadronic scale.
Those three CP-odd operators are known to comparably contribute to the nucleon EDM, and this fact requires tremendous efforts in the QCD calculation.
The nuclear EDM is providing an alternative approach in the determination of the Weinberg operator, together with its own difficulty.
The Weinberg operator can be probed through the isoscalar contact CP-odd nucleon-nucleon interaction, but the calculation of this process suffers from a large uncertainty in the nuclear level calculation, due to the lack of informations of the nuclear force at short distance.
This problem may be resolved if the short range behavior of the CP-even nuclear force is determined in high energy experiments \cite{sargsian,miller}, or by lattice QCD calculations \cite{hal1,hal2}.
There are also no available results for the QCD calculation of the Weinberg operator contribution to the contact CP-odd nucleon-nucleon interaction.
Moreover, the contribution of the quark chromo-EDM appearing from the mixing due to the renormalization group evolution also has to be quantified.
Those issues can only be resolved using lattice QCD, if we wish to determine it beyond the order estimation.

\subsection{Nuclear level enhancement of CP violation}

Let us also see the prospects for finding nuclei sensitive on the nucleon level CP violation.
In Section \ref{sec:edm6li9be13c}, we have seen that the bad transition between the ground state and low lying opposite parity states with the same angular momentum suppresses the polarization contribution to the nuclear EDM.
Light nuclei with a dominantly shell-like structure have this tendency.
We therefore have to find nuclei with a well developed cluster structure.
Restricting to stable nuclei, good candidates are $^7$Li and $^{19}$F.

The $^7$Li is known to have a $\alpha - ^3$H cluster structure.
The $\alpha - ^3$H threshold is just 2.5 MeV above the ground state, and the CP-odd transition to the continuum is expected to be large, as the $\alpha$ core does not change its structure.
The $^7$Li nucleus is manageable in the four-body cluster model ($\alpha - p -p -n$), and it is thus a good target of future work.
Although being unstable, $^7$Be, the mirror nucleus of $^7$Li, may also be a good candidate since it has a life time of 53 days, which is experimentally manageable.
In this case, the Coulomb force decreases the energy needed to reach the $\alpha- ^3$He threshold from the ground state to 1.6 MeV, so that parity violation is enhanced, and the EDM may be increased.

The $^{19}$F nucleus is also a good candidate, as it can be described by the mixing of $^{15}$N$-\alpha$ and $^{16}$O$-^{3}$H cluster states.
The remarkable point is that its first opposite parity excited state ($\frac{1}{2}^-$) is just 109 keV above the ground state ($\frac{1}{2}^+$), so that a sizable enhancement of the nuclear polarization by the isovector CP-odd nuclear force is expected.
This nucleus can be calculated in the two-body cluster model with coupled channel OCM \cite{sakuda1,sakuda2}.

\section{Summary\label{sec:summary}}

In this review, we have summarized the current status of the theoretical development of the EDM of light nuclei.
The EDM of $^2$H, $^3$He, $^3$H, $^6$Li, $^9$Be, and $^{13}$C nuclei have so far been evaluated.
The study of the nuclear polarization of those nuclei by CP-odd effects is showing that the cluster structure may enhance the sensitivity on nucleon level CP violation, as for the case of $^6$Li.
Moreover, this extensive analysis is also suggesting the existence of nuclei furthermore sensitive on CP violation, such as $^7$Li, $^7$Be or $^{19}$F, and their theoretical investigations are strongly wanted.
The study of those nuclei will be an important target of our future work.

The EDM of light nuclei have dependences on various parameters of new physics BSM. 
Among them, the most important nucleon level CP violation for light nuclei is the isovector CP-odd nuclear force.
However, as the determination of the new physics BSM is not achievable by only measuring the EDM of one very sensitive system, the combination of several experimental data of different observables is mandatory in constraining the large parameter space of new physics.
To discern the CP violation BSM, the combined analysis with the experimental search of the single nucleon EDM, the atomic EDM, and T-odd neutron-nucleus scattering may be efficient, as these observables are sensitive to the isoscalar CP-odd pion-nucleon interaction.

The search for CP violation is not restricted to the analysis of the EDM.
Recently there are also efforts in constraining the TeV scale CP violation through the combined analysis with the experimental data of the LHC experiments \cite{inoue3,dekens3}, or with the simulation of the baryogenesis which occurred in the early universe \cite{fuyuto2,Kobakhidze,Balazs}.
The combination of the EDM search with other approaches is also important in the determination of the origin of matter of our Universe.

The experimental measurements of the EDM of the proton and light nuclei are prepared at FermiLab and at J\"{u}lich, with a prospective experimental sensitivity of $d_A \sim O(10^{-29})e$ cm.
The theoretical uncertainty due to the QCD calculation of low energy constants is still large, and further investigations are needed to fix at least the first digit, to be confident in observing signals BSM.
The prospective experimental sensitivity has the potential to unveil typical models of new physics BSM between the energy scale of TeV to PeV.
By experimentally observing the EDM of the deuteron or three-nucleon systems, significant advance in the understanding of new physics is expected.
We also recommend the development of the experimental and theoretical studies of the EDM of other heavier light nuclei.

\section*{Acknowledgments}

The author is very grateful to Jordy de Vries who kindly helped him improving the review.
He also thanks the organizers and participants of the Workshop "Baryons over antibaryons: the nuclear physics of Sakharov", held at ECT* - European Centre for Theoretical Studies in Nuclear Physics and Related Areas, Italy,  in July 2016, for useful discussions and comments.
This work was completed due to support of the RSF grant  15-12-20008.
It was also supported by Riken iTHES project.



\begin{thebibliography}{0}    

\bibitem{planck}
P. A. R. Ade {\it et al.} (Planck Collaboration), Astron. Astrophys. {\bf 571}, 66 (2014).

\bibitem{sakharov}
A. D. Sakharov, Pisma Zh. Eksp. Teor. Fiz. {\bf 5}, 32 (1967) [JETP Lett. {\bf 5}, 24 (1967)].

\bibitem{farrar}
G. R. Farrar and M. E. Shaposhnikov, Phys. Rev. D {\bf 50}, 774 (1994).

\bibitem{huet}
P. Huet and E. Sather, Phys. Rev. D {\bf 51}, 379 (1995).

\bibitem{ckm}
M. Kobayashi and T. Maskawa, Prog. Theor. Phys. {\bf 49}, 652 (1973).

\bibitem{jarlskog}
C. Jarlskog, Phys. Rev. Lett. {\bf 55}, 1039 (1985).

\bibitem{cronin}
J. H. Christenson, J. W. Cronin, V. L. Fitch and R. Turlay, Phys. Rev. Lett. {\bf 13}, 138 (1964).

\bibitem{hereview}
X.-G. He, B. H. J. McKellar and S. Pakvasa, Int. J. Mod. Phys. A {\bf 4}, 5011 (1989) [Erratum-ibid. A {\bf 6}, 1063 (1991)].

\bibitem{bernreuther}
W. Bernreuther and M. Suzuki, Rev. Mod. Phys. {\bf 63}, 313 (1991); Erratum-ibid. {\bf 64}, 633 (1992).

\bibitem{khriplovichbook}
I. B. Khriplovich and S. K. Lamoreaux, CP Violation Without Strangeness (Springer, Berlin, 1997).

\bibitem{ginges}
J. S. M. Ginges and V. V. Flambaum, Phys. Rept. {\bf 397}, 63 (2004).

\bibitem{pospelovreview}
M. Pospelov and A. Ritz, Ann. Phys. {\bf 318}, 119 (2005).

\bibitem{fukuyamareview}
T. Fukuyama, Int. J. Mod. Phys. A {\bf 27}, 1230015 (2012).

\bibitem{hewett}
J. L. Hewett {\it et al.}, arXiv:1205.2671 [hep-ex].

\bibitem{engelreview}
J. Engel, M. J. Ramsey-Musolf and U. van Kolck, Prog. Part. Nucl. Phys. {\bf 71}, 21 (2013).

\bibitem{yamanakabook}
N. Yamanaka, ``Analysis of the Electric Dipole Moment in the R-parity Violating Supersymmetric Standard Model'', Springer, Berlin, Germany (2014).

\bibitem{devriesreview}
J. de Vries and Ulf-G. Mei{\ss}ner, Int. J. Mod. Phys. E {\bf 25}, 1641008 (2016).

\bibitem{purcell}
E. M. Purcell and N. F. Ramsey, Phys. Rev. {\bf 78}, 807 (1950).

\bibitem{baker}
C. A. Baker {\it et al}., 
Phys. Rev. Lett. {\bf 97}, 131801 (2006).

\bibitem{rosenberry}
M. A. Rosenberry and T. E. Chupp, 
Phys. Rev. Lett. {\bf 86}, 22 (2001).

\bibitem{regan}
B.C. Regan, E.D. Commins, C.J. Schmidt and D. DeMille, 
Phys. Rev. Lett. {\bf 88}, 071805 (2002).

\bibitem{griffith}
W. C. Griffith {\it et al}., 
Phys. Rev. Lett. {\bf 102}, 101601 (2009).

\bibitem{hudson}
J. J. Hudson {\it et al.}, 
Nature {\bf 473}, 493 (2011).

\bibitem{acme}
J. Baron {\it et al}. (ACME collaboration), 
Science {\bf 343}, 269 (2014). 

\bibitem{parker}
R. H. Parker {\it et al.}, 
Phys. Rev. Lett. {\bf 114}, 233002 (2015). 

\bibitem{graner}
B. Graner, Y. Chen, E. G. Lindahl and B. R. Heckel, 
Phys. Rev. Lett. {\bf 116}, 161601 (2016). 

\bibitem{bishof}
M. Bishof {\it et al.}, 
Phys. Rev. C {\bf 94}, 025501 (2016). 

\bibitem{muong2}
G. W. Bennett {\it et al.} (Muon ($g-2$) Collaboration), Phys. Rev. D {\bf 80}, 052008 (2009).

\bibitem{storage1}
I. B. Khriplovich, Phys. Lett. B {\bf 444}, 98 (1998).

\bibitem{storage2}
F. J. M. Farley {\it et al.}, Phys. Rev. Lett. {\bf 93}, 052001 (2004).

\bibitem{storage3}
Y. K. Semertzidis {\it et al.}, AIP Conf. Proc. {\bf 698}, 200 (2004).

\bibitem{storage4}
Y. F. Orlov, W. M. Morse, and Y. K. Semertzidis, Phys. Rev. Lett. {\bf 96}, 214802 (2006).

\bibitem{storage5}
A. Lehrach, B. Lorentz, W. Morse, N. Nikolaev and F. Rathmann, arXiv:1201.5773 [hep-ex].

\bibitem{storage6}
T. Fukuyama and A. J. Silenko, Int. J. Mod. Phys. A {\bf 28}, 1350147 (2013). 

\bibitem{storage7}
T. Fukuyama, Mod. Phys. Lett. A {\bf 31}, 1650135 (2016). 

\bibitem{storage8}
V. Anastassopoulos {\it et al.}, Rev. Sci. Instrum. {\bf 87}, 115116 (2016). 

\bibitem{storage9}
D. Eversmann {\it et al.} (JEDI Collaboration), Phys. Rev. Lett. {\bf 115}, 094801 (2015). 

\bibitem{storage10}
G. Guidoboni {\it et al.} (JEDI Collaboration), 
Phys. Rev. Lett. {\bf 117}, 054801 (2016).

\bibitem{bnl}
Storage Ring EDM Collaboration, http://www.bnl.gov/edm/.

\bibitem{schiff}
L. I. Schiff, Phys. Rev. {\bf 132}, 2194 (1963).

\bibitem{sushkov}
O. P. Sushkov, V. V. Flambaum and I. B. Khriplovich, 
Zh. Eksp. Teor. Fiz. {\bf 87}, 1521 (1984) [Sov. Phys. JETP {\bf 60}, 873 (1984)].

\bibitem{sushkov2}
V. V. Flambaum, O. P. Sushkov and I. B. Khriplovich, 
Phys. Lett. B {\bf 162} 213 (1985).

\bibitem{sushkov3}
V. V. Flambaum, O. P. Sushkov and I. B. Khriplovich, 
Nucl. Phys. A {\bf 449}, 750 (1986).

\bibitem{chiral3nucleon}
J. de Vries, R. Higa, C.-P. Liu, E. Mereghetti, I. Stetcu, R. G. E. Timmermans and U. van Kolck, Phys. Rev. C {\bf 84}, 065501 (2011).

\bibitem{mereghetti2}
E. Mereghetti, W. H. Hockings and U. van Kolck, Annals Phys. {\bf 325}, 2363 (2010).

\bibitem{crewther}
R. J. Crewther, P. Di Vecchia, G. Veneziano and E. Witten, 
Phys. Lett. B {\bf 88}, 123 (1979) [Erratum ibid. B {\bf 91}, 487 (1980)].

\bibitem{pich}
A. Pich and E. de Rafael, Nucl. Phys. B {\bf 367}, 313 (1991).

\bibitem{borasoy}
B. Borasoy, Phys. Rev. D {\bf 61}, 114017 (2000).

\bibitem{narison}
S. Narison, Phys. Lett. B {\bf 666}, 455 (2008).

\bibitem{ottnad}
K. Ottnad, B. Kubis, U.-G. Mei{\ss}ner and F.-K. Guo, Phys. Lett. B {\bf 687}, 42 (2010).

\bibitem{mereghetti1}
E. Mereghetti, J. de Vries, W. H. Hockings, C. M. Maekawa and U. van Kolck, Phys. Lett. B {\bf 696}, 97 (2011).

\bibitem{devries1}
J. de Vries, R. G. E. Timmermans, E. Mereghetti and U. van Kolck, Phys. Lett. B {\bf 695}, 268 (2011).

\bibitem{guo1}
F.-K. Guo and U.-G. Mei{\ss}ner, JHEP {\bf 1212}, 097 (2012).

\bibitem{ucna}
B. Plaster {\it et al.} (UCNA Collaboration), 
Phys. Rev. C {\bf 86}, 055501 (2012). 

\bibitem{fuyuto}
K. Fuyuto, J. Hisano and N. Nagata, 
Phys. Rev. D {\bf 87}, 054018 (2013). 

\bibitem{maekawa}
C. M. Maekawa, E. Mereghetti, J. de Vries, and U. van Kolck, Nucl. Phys. A {\bf 872}, 117 (2011). 

\bibitem{Barton}
G. Barton and E. G.White, Phys. Rev. {\bf 184}, 1660 (1969).

\bibitem{pvcpvhamiltonian1}
W. C. Haxton and E. M. Henley, {\it Phys. Rev. Lett.} {\bf 51}, 1937 (1983).

\bibitem{pvcpvhamiltonian2}
V. P. Gudkov, X.-G. He and B. H. J. McKellar, {\it Phys. Rev. C} {\bf 47}, 2365 (1993).

\bibitem{pvcpvhamiltonian3}
I. S. Towner and A. C. Hayes, {\it Phys. Rev. C} {\bf 49}, 2391 (1994).

\bibitem{korkin}
I. B. Khriplovich and R. A. Korkin, Nucl. Phys. A {\bf 665}, 365 (2000).

\bibitem{liu}
C.-P. Liu and R. G. E. Timmermans, Phys. Rev. C {\bf 70}, 055501 (2004).

\bibitem{stetcu}
I. Stetcu, C.-P. Liu, J. L. Friar, A. C. Hayes and P. Navratil, Phys. Lett. B {\bf 665}, 168 (2008).

\bibitem{afnan}
I. R. Afnan and B. F. Gibson, Phys. Rev. C {\bf 82}, 064002 (2010).

\bibitem{song}
Y.-H. Song, R. Lazauskas and V. Gudkov, {\it Phys. Rev. C} {\bf 87}, 015501 (2013).

\bibitem{yamanakanuclearedm}
N. Yamanaka and E. Hiyama, Phys. Rev. C {\bf 91}, 054005 (2015).

\bibitem{auerbach}
N. Auerbach, V. V. Flambaum and V. Spevak, Phys. Rev. Lett. {\bf 76}, 4316 (1996).

\bibitem{Spevak}
V. Spevak, N. Auerbach and V. V. Flambaum, Phys. Rev. C {\bf 56}, 1357 (1997).

\bibitem{dobaczewski}
J. Dobaczewski and J. Engel, 
Phys. Rev. Lett. {\bf 94}, 232502 (2005). 

\bibitem{eft6dim}
J. de Vries, E. Mereghetti, R.G.E. Timmermans and U. van Kolck, 
Annals Phys. {\bf 338}, 50 (2013). 

\bibitem{av18}
R. B. Wiringa, V. G. J. Stoks and R. Schiavilla, Phys. Rev. C {\bf 51}, 38 (1995).

\bibitem{bsaisou}
J. Bsaisou, J. de Vries, C. Hanhart, S. Liebig, U.-G. Mei{\ss}ner, D. Minossi, A. Nogga and A. Wirzba, J. High Energy Phys. {\bf 1503}, 104 (2015) [Erratum ibid. {\bf 1505}, 083 (2015)]. 

\bibitem{bsaisou2}
J. Bsaisou, U.-G. Mei{\ss}ner, A. Nogga and A. Wirzba, Annals Phys. {\bf 359}, 317 (2015).

\bibitem{13cedm}
N. Yamanaka, T. Yamada, E. Hiyama and Y. Funaki, arXiv:1603.03136 [nucl-th].

\bibitem{sandars1}
P. G. H. Sandars, Phys. Lett. {\bf 14}, 194 (1965).

\bibitem{sandars2}
P. G. H. Sandars, Phys. Lett. {\bf 22}, 290 (1966).

\bibitem{flambaum2}
V. V. Flambaum, Yad. Fiz. {\bf 24}, 383 (1976) [Sov. J. Nucl. Phys. {\bf 24}, 199 (1976)].

\bibitem{sinoue}
S. Inoue, V. Gudkov, M. R. Schindler and Y.-H. Song, Phys. Rev. C {\bf 93}, 055501 (2016). 

\bibitem{yoshinaga1}
N. Yoshinaga, K. Higashiyama and R. Arai, Prog. Theor. Phys. {\bf 124}, 1115 (2010).

\bibitem{yoshinaga2}
N. Yoshinaga, K. Higashiyama, R. Arai and E. Teruya, Phys. Rev. C {\bf 89}, 045501 (2014).

\bibitem{fujita}
T. Fujita and S. Oshima, J. Phys. G {\bf 39}, 095106 (2012).

\bibitem{nijmegen}
V. G. J. Stoks, R. A. M. Klomp, C. P. F. Terheggen and J. J. de Swart, Phys. Rev. C {\bf 49}, 2950 (1994).

\bibitem{inoy}
P. Doleschall, I. Borbely, Z. Papp and W. Plessas, Phys. Rev. C {\bf 67}, 064005 (2003).

\bibitem{3bodyforce1}
J. Carlson, V. R. Pandharipande, and R. B. Wiringa, Nucl. Phys. A {\bf 401}, 59 (1983).

\bibitem{3bodyforce2}
B. S. Pudliner, V. R. Pandharipande, J. Carlson and R. B. Wiringa, Phys. Rev. Lett. {\bf 74}, 4396 (1995).

\bibitem{3bodyforce3}
R. B. Wiringa and S. C. Pieper, Phys. Rev. Lett. {\bf 89}, 182501 (2002).

\bibitem{3bodyforce4}
R. B. Wiring, R. Schiavilla, S. C. Pieper and J. Carlson, Phys. Rev. C {\bf 89}, 024305 (2014).

\bibitem{clusterreview1}
W. von Oertzen, M. Freer and Y. Kanada-En'yo, 
Phys. Rep. {\bf 432}, 43 (2006).

\bibitem{clusterreview2}
H. Horiuchi, K. Ikeda and K. Kato, 
Prog. Theor. Phys. Suppl. {\bf 192}, 1 (2011).

\bibitem{clusterreview3}
Y. Funaki, H. Horiuchi and A. Tohsaki, 
Prog. Part. Nucl. Phys. {\bf 82}, 78 (2015).

\bibitem{schmid}
E. W. Schmid and K. Wildermuth, 
Nucl. Phys. {\bf 26}, 463 (1961).

\bibitem{hasegawa}
A. Hasegawa and S. Nagata, Prog. Theor. Phys. {\bf 45}, 1786 (1971).

\bibitem{kanada}
H. Kanada, T. Kaneko, S. Nagata and M. Morikazu, {\it Prog. Theor. Phys.} {\bf 61}, 1327 (1979).

\bibitem{ocm1}
S. Saito, 
Prog. Theor. Phys. {\bf 40}, 893 (1968).

\bibitem{ocm2}
S. Saito, 
Prog. Theor. Phys. {\bf 41}, 705 (1969).

\bibitem{horiuchi1}
H. Horiuchi, 
Prog. Theor. Phys. {\bf 58}, 204 (1977).

\bibitem{ocm3}
S. Saito, 
Prog. Theor. Phys. Suppl. No. {\bf 62}, 11 (1977).

\bibitem{horiuchi2}
H. Horiuchi, 
Prog. Theor. Phys. Suppl. {\bf 62}, 90 (1977).

\bibitem{Kukulin}
V. I. Kukulin, V. N. Pomerantsev, Kh. D. Razikov, V. T. Voronchev and G. G. Ryzhinkh, 
Nucl. Phys. A {\bf 586}, 151 (1995).

\bibitem{funaki8Be}
Y. Funaki, H. Horiuchi, A. Tohsaki, P. Schuck and G. R\"{o}pke, Prog. Theor. Phys. {\bf 108}, 297 (2002).

\bibitem{funaki12C}
Y. Funaki, A. Tohsaki, H. Horiuchi, P. Schuck and G. Röpke, Phys. Rev. C {\bf 67}, 051306 (2003).

\bibitem{funaki16O}
Y. Funaki, T. Yamada, H. Horiuchi, G. Röpke, P. Schuck and A. Tohsaki, Phys. Rev. Lett. {\bf 101}, 082502 (2008).

\bibitem{yamada13C}
T. Yamada and Y. Funaki, 
Phys. Rev. C {\bf 92}, 034326 (2015). 

\bibitem{hiyama}
E. Hiyama, Y. Kino and M. Kamimura, 
Prog. Part. Nucl. Phys. {\bf 51}, 223 (2003).

\bibitem{pospelovtheta1}
M. Pospelov and A. Ritz, 
Phys. Rev. Lett. {\bf 83}, 2526 (1999).

\bibitem{pospelovtheta2}
M. Pospelov and A. Ritz, 
Nucl. Phys. B {\bf 573}, 177 (2000).

\bibitem{nedmholography}
D. K. Hong, H.-C. Kim, S. Siwach and H.-U. Yee, 
JHEP {\bf 0711}, 036 (2007). 

\bibitem{shintani2}
E. Shintani, S. Aoki and Y. Kuramashi, Phys. Rev. D {\bf 78}, 014503 (2008).

\bibitem{nedmlattice1}
S. Aoki, A. Gocksch, A. V. Manohar and Stephen R. Sharpe, Phys. Rev. Lett. {\bf 65}, 1092 (1990).

\bibitem{shintani1}
E. Shintani {\it et al.}, Phys. Rev. D {\bf 72}, 014504 (2005).

\bibitem{nedmlattice3}
F. Berruto, T. Blum, K. Orginos and A. Soni, Phys. Rev. D {\bf 73}, 054509 (2006).

\bibitem{nedmlattice5}
F.-K. Guo, R. Horsley, U.-G. Meissner, Y. Nakamura, H. Perlt, P. E. L. Rakow, G. Schierholz, A. Schiller and J. M. Zanotti, 
Phys. Rev. Lett. {\bf 115}, 062001 (2015). 

\bibitem{nedmlattice6}
A. Shindler, T. Luu and J. de Vries, 
Phys. Rev. D {\bf 92}, 094518 (2015). 

\bibitem{nedmetm}
C. Alexandrou, A. Athenodorou, M. Constantinou, K. Hadjiyiannakou, K. Jansen, G. Koutsou, K. Ottnad and M. Petschlies, 
Phys. Rev. D {\bf 93}, 074503 (2016). 

\bibitem{shintani3}
E. Shintani, T. Blum, T. Izubuchi and A. Soni, 
Phys. Rev. D {\bf 93}, 094503 (2016). 

\bibitem{peccei}
R. D. Peccei and H. R. Quinn, 
Phys. Rev. Lett. {\bf 38}, 1440 (1977).

\bibitem{bigi1}
I. Bigi and N. G. Ural'tsev, Zh. Eksp. Teor. Fiz. {\bf 100}, 363 (1991) [Sov. Phys. JETP {\bf 73}, 198 (1991)].

\bibitem{bigi2}
I. Bigi and N. G. Ural'tsev, Nucl. Phys. B {\bf 353}, 321 (1991).

\bibitem{belayev1}
V. M. Belayev and I. B. Ioffe, Sov. Phys. JETP {\bf 100}, 493 (1982).

\bibitem{belayev2}
V. M. Belayev and Ya. I. Kogan, Sov. J. Nucl. Phys.  {\bf 40}, 659 (1984).

\bibitem{devriesdeuteron}
J. de Vries, E. Mereghetti, R. G. E. Timmermans, and U. van Kolck, Phys. Rev. Lett. {\bf 107}, 091804 (2011).

\bibitem{bsaisoudeuterontheta}
J.~Bsaisou, C.~Hanhart, S.~Liebig, U.-G.~Meissner, A.~Nogga and A.~Wirzba, Eur. Phys. J. A {\bf 49}, 31 (2013).

\bibitem{mereghettitheta}
E. Mereghetti and U. van Kolck, Ann. Rev. Nucl. Part. Sci. {\bf 65}, 215 (2015).

\bibitem{devriessplitting}
J. de Vries, E. Mereghetti and A. Walker-Loud, 
Phys. Rev. C {\bf 92}, 045201 (2015). 

\bibitem{pdg}
K. A. Olive {\it et al.} (Particle Data Group), Chin. Phys. C, {\bf 38} 090001 (2014). 

\bibitem{avishai1}
Y. Avishai, 
Phys. Rev. D {\bf 32}, 314 (1985).

\bibitem{avishai2}
Y. Avishai and M. Fabre de la Ripelle, 
Phys. Rev. Lett. {\bf 56}, 2121 (1986).

\bibitem{avishai3}
Y. Avishai and M. Fabre de la Ripelle, 
Nucl. Phys. A {\bf 468}, 578 (1987).

\bibitem{donoghue1}
J. F. Donoghue, B. R. Holstein and M. J. Musolf, Phys. Lett. B {\bf 196}, 196 (1987).

\bibitem{hesmedm}
X.-G. He and B. McKellar, Phys. Rev. D {\bf 46}, 2131 (1992).

\bibitem{ellisthetasm}
J. Ellis and M. K. Gaillard, Nucl. Phys. B {\bf 150}, 141 (1979).

\bibitem{khriplovichtheta}
I. B. Khriplovich, Phys. Lett. B {\bf 173}, 193 (1986).

\bibitem{gerard}
J.-M. G\'{e}rard and P. Mertens, Phys. Lett. B {\bf 716}, 316 (2012).

\bibitem{czarnecki}
A. Czarnecki and B. Krause, Phys. Rev. Lett. {\bf 78}, 4339 (1997).

\bibitem{smweinbergop}
M. E. Pospelov, Phys. Lett. B {\bf 328}, 441 (1994).

\bibitem{yamanakasmedm}
N. Yamanka and E. Hiyama, J. High Energy Phys. {\bf 02}, 067 (2016).

\bibitem{buras}
A. J. Buras, M. Jamin, M. E. Lautenbacher and P. H. Weisz, Nucl. Phys. B {\bf 370}, 69 (1992) [Erratum ibid. B {\bf 375}, 501 (1992)].

\bibitem{buras2}
G. Buchalla, A. J. Buras and M. E. Lautenbacher, Rev. Mod. Phys. {\bf 68}, 1125 (1996). 

\bibitem{smdeuteronedm}
N. Yamanaka, Nucl. Phys. A {\bf 963}, 33 (2017). 

\bibitem{donoghue2}
J. F. Donoghue and S. Pakvasa, Phys. Rev. Lett. {\bf 55}, 162 (1985).

\bibitem{donoghue3}
J. F. Donoghue, X.-G. He and S. Pakvasa, Phys. Rev. D {\bf 34}, 833 (1986).

\bibitem{he1}
X. G. He, H. Steger and G. Valencia, Phys. Lett. B {\bf 272}, 411 (1991).

\bibitem{hiyamahyperondecay}
E. Hiyama, K. Suzuki, H. Toki, and M. Kamimura, Prog. Theor. Phys. {\bf 112}, 99 (2004). 

\bibitem{inoue1}
T. Inoue, S. Takeuchi and M. Oka, Nucl. Phys. A {\bf 597}, 563 (1996). 

\bibitem{inoue2}
T. Inoue, M. Oka, T. Motoba and K. Itonaga, Nucl. Phys. A {\bf 633}, 312 (1998). 

\bibitem{sasaki1}
K. Sasaki, T. Inoue and M. Oka, Nucl. Phys. A {\bf 669}, 331 (2000) [Erratum ibid. Nucl. Phys. A {\bf 678}, 455 (2000)]. 

\bibitem{sasaki2}
K. Sasaki, T. Inoue and M. Oka, Nucl. Phys. A {\bf 707}, 477 (2002). 

\bibitem{perez1}
A. P\'{e}rez-Obiol, A. Parre\~{n}o and B. Juli\'{a}-D\'{i}az, Phys. Rev. C {\bf 84}, 024606 (2011).

\bibitem{perez2}
A. P\'{e}rez-Obiol, D. R. Entem, B. Juli\'{a}-D\'{i}az and A. Parre\~{n}o, Phys. Rev. C {\bf 87}, 044614 (2013).

\bibitem{perez3}
A. P\'{e}rez-Obiol, D. R. Entem, B. Juli\'{a}-D\'{i}az and A. Parre\~{n}o, Nucl. Phys. A {\bf 954}, 213 (2016). 

\bibitem{ellis}
J. Ellis, M. K. Gaillard and D. V. Nanopoulos, Nucl. Phys. B {\bf 109}, 213 (1976).

\bibitem{khriplovichnedm}
I. B. Khriplovich and A. R. Zhitnitsky, Phys. Lett. B {\bf 109}, 490 (1982).

\bibitem{mckellar1}
B. H. J. McKellar, S. R. Choudhury, X.-G. He and S. Pakvasa, Phys. Lett. B {\bf 197}, 556 (1987).

\bibitem{seng}
C.-Y. Seng, Phys. Rev. C {\bf 91}, 025502 (2015).

\bibitem{mannel}
T. Mannel and N. Uraltsev, Phys. Rev. D {\bf 85}, 096002 (2012).

\bibitem{qcdsf1}
G. S. Bali {\it et al.} (QCDSF Collaboration), 
Phys. Rev. D {\bf 85}, 054502 (2012).

\bibitem{engelhardt}
M. Engelhardt, 
Phys. Rev. D {\bf 86}, 114510 (2012). 

\bibitem{jlqcd2}
K. Takeda, S. Aoki, S. Hashimoto, T. Kaneko, J. Noaki and T. Onogi (JLQCD Collaboration), Phys. Rev. D {\bf 83}, 114506 (2011).

\bibitem{jlqcd3}
H. Ohki, K. Takeda, S. Aoki, S. Hashimoto, T. Kaneko, H. Matsufuru, J. Noaki and T. Onogi (JLQCD Collaboration), Phys. Rev. D {\bf 87}, 034509 (2013). 

\bibitem{junnarkar}
P. M. Junnarkar and A. Walker-Loud, 
Phys. Rev. D {\bf 87}, 114510 (2013). 

\bibitem{gong}
M. Gong {\it et al.}, 
Phys. Rev. D {\bf 88}, 014503 (2013). 

\bibitem{etm1}
C. Alexandrou, M. Constantinou, S. Dinter, V. Drach, K. Hadjiyiannakou, K. Jansen, G. Koutsou and A. Vaquero, 
Phys. Rev. D {\bf 91}, 094503 (2015). 

\bibitem{etm2}
A. Abdel-Rehim {\it et al.}, 
Phys. Rev. D {\bf 89}, 034501 (2014).

\bibitem{chiqcdsigmaterm}
Y.-B. Yang, A. Alexandru, T. Draper, J. Liang and K.-F. Liu ($\chi$QCD Collaboration), Phys. Rev. D {\bf 94}, 054503 (2016).

\bibitem{rqcdsigmaterm}
G. S. Bali {\it et al.}, Phys. Rev. D {\bf 93}, 094504 (2016).

\bibitem{bmwsigmaterm}
S. Durr {\it et al.}, 
Phys. Rev. Lett. {\bf 116}, 172001 (2016). 

\bibitem{etmsigmaterm}
A. Abdel-Rehim {\it et al.}, Phys. Rev. Lett. {\bf 116}, 252001 (2016).

\bibitem{jlqcd4}
N. Yamanaka, H. Ohki, S. Hashimoto and T. Kaneko (JLQCD Collaboration),  PoS LATTICE2015 (2016) 121. 

\bibitem{mssmreloaded}
J. R. Ellis, J. S. Lee and A. Pilaftsis, 
JHEP {\bf 0810}, 049 (2008). 

\bibitem{ellismssmedm}
J. R. Ellis, S. Ferrera and D. V. Nanopoulos, Phys. Lett. B {\bf 114}, 231 (1982).

\bibitem{demir}
D. Demir, O. Lebedev, K. A. Olive, M. Pospelov and A. Ritz, Nucl. Phys. B {\bf 680}, 339 (2004).

\bibitem{xu}
F. Xu, H. An and X. Ji, JHEP {\bf 1003}, 088 (2010).

\bibitem{nemev}
A. Maiezza and M. Nemev\u{s}ek, Phys. Rev. D {\bf 90}, 095002 (2014).

\bibitem{dekens2}
W.~Dekens, J.~de Vries, J.~Bsaisou, W.~Bernreuther, C.~Hanhart, U.~G.~Mei{\ss}ner, A.~Nogga and A.~Wirzba, 
J. High Energy Phys. {\bf 1407}, 069 (2014).

\bibitem{faessler}
A. Faessler, T. Gutsche, S. Kovalenko and V. E. Lyubovitskij, Phys. Rev. D {\bf 74}, 074013 (2006).

\bibitem{2higgs}
S. M. Barr and A. Zee, Phys. Rev. Lett. {\bf 65}, 21 (1990).

\bibitem{chang2hdm1}
D. Chang, W.-Y. Keung and T. C. Yuan, Phys. Lett. B {\bf 251}, 608 (1990).

\bibitem{abe}
T. Abe, J. Hisano, T. Kitahara and K. Tobioka, J. High Energy Phys. {\bf 1401}, 106 (2014).

\bibitem{chang}
D. Chang, W.-Y. Keung and A. Pilaftsis, Phys. Rev. Lett. {\bf 82}, 900 (1999).

\bibitem{pilaftsis}
A. Pilaftsis, 
Phys. Lett. B {\bf 471}, 174 (1999).

\bibitem{chang2}
D. Chang, W.-F. Chang and W.-Y. Keung, 
Phys. Lett. B {\bf 478}, 239 (2000).

\bibitem{yamanakarainbow1}
N. Yamanaka, Phys. Rev. D {\bf 87}, 011701 (2013).

\bibitem{rpvyamanaka1}
N. Yamanaka, T. Sato and T. Kubota, Phys. Rev. D {\bf 85}, 117701 (2012).

\bibitem{rpvyamanaka2}
N. Yamanaka, Phys. Rev. D {\bf 86}, 075029 (2012).

\bibitem{rpvyamanaka3}
N. Yamanaka, T. Sato and T. Kubota, Phys. Rev. D {\bf 87}, 115011 (2013).

\bibitem{arkani-hamed}
N. Arkani-Hamed, S. Dimopoulos, G. F. Giudice and A. Romanino, 
Nucl. Phys. B {\bf 709}, 3 (2005).

\bibitem{chang3}
D. Chang, W.-F. Chang and W.-Y. Keung, Phys. Rev. D {\bf 71}, 076006 (2005).

\bibitem{giudice}
G. F. Giudice and A. Romanino, 
Phys. Lett. B {\bf 634}, 307 (2006).

\bibitem{dhuria}
M. Dhuria and A. Misra, Phys. Rev. D {\bf 90}, 085023 (2014).

\bibitem{bowser-chao}
D. Bowser-Chao, D. Chang and W.-Y. Keung, 
Phys. Rev. Lett. {\bf 79}, 1988 (1997). 

\bibitem{tensorrenormalization1}
X. Artru and M. Mekhfi, Z. Phys. C {\bf 45}, 669 (1990).

\bibitem{tensorrenormalization2}
V. Barone, Phys. Lett. B {\bf 409}, 499 (1997).

\bibitem{degrassi}
G. Degrassi and S. Marchetti, E. Franco and L. Silvestrini, 
JHEP {\bf 0511}, 044 (2005). 

\bibitem{yang}
J. Hisano, K. Tsumura and M. J. S. Yang, 
Phys. Lett. B {\bf 713}, 473 (2012). 

\bibitem{dekens1}
W. Dekens and J. de Vries, J. High Energy Phys. {\bf 1305}, 149 (2013).

\bibitem{yamanakasde1}
N. Yamanaka, T. M. Doi, S. Imai and H. Suganuma, Phys. Rev. D {\bf 88}, 074036 (2013).

\bibitem{pitschmann}
M. Pitschmann, C.-Y. Seng, C. D. Roberts, and S. M. Schmidt, Phys. Rev. D {\bf 91}, 074004 (2015). 

\bibitem{radici}
M. Radici, A. Courtoy, A. Bacchetta and M. Guagnelli, J. High Energy Phys. {\bf 1505}, 123 (2015). 

\bibitem{bacchetta}
A. Bacchetta, A. Courtoy and M. Radici, J. High Energy Phys. {\bf 1303}, 119 (2013).

\bibitem{courtoy}
A. Courtoy, S. Bae{\ss}ler, M. Gonz\'{a}lez-Alonso and S. Liuti, Phys. Rev. Lett. {\bf 115}, 162001 (2015).

\bibitem{kang}
Z.-B. Kang, A. Prokudin, P. Sun and F. Yuan, Phys. Rev. D {\bf 93}, 014009 (2016).

\bibitem{gutsche}
T. Gutsche, M. A. Ivanov, J. G. Korner, S. Kovalenko and V. E. Lyubovitskij, Phys. Rev. D {\bf 94}, 114030 (2016).

\bibitem{yez}
Z. Ye {\it et al.}, Phys. Lett. B {\bf 767}, 91 (2017). 

\bibitem{green}
J. R. Green, J. W. Negele, A. V. Pochinsky, S. N. Syritsyn, M. Engelhardt and S. Krieg, Phys. Rev. D {\bf 86}, 114509 (2012).

\bibitem{rbcukqcdisovectortensor}
Y. Aoki, T. Blum, H.-W. Lin, S. Ohta, S. Sasaki, R. Tweedie, J. Zanotti and T. Yamazaki, Phys. Rev. D {\bf 82}, 014501 (2010).

\bibitem{rqcdisovector}
G. S. Bali {\it et al.} (RQCD Collaboration), Phys. Rev. D {\bf 91}, 054501 (2015).

\bibitem{chiqcdisovector}
Y.-B. Yang, A. Alexandru, T. Draper, M. Gong, and K.-F. Liu, Phys. Rev. D {\bf 93}, 034503 (2016).

\bibitem{pndmeisovector}
T. Bhattacharya {\it et al.} (PNDME Collaboration), Phys. Rev. D {\bf 94}, 054508 (2016). 

\bibitem{etmisovector}
A. Abdel-Rehim {\it et al.}, Phys. Rev. D {\bf 92}, 114513 (2015).

\bibitem{pndmetensor1}
T. Bhattacharya, V. Cirigliano, R. Gupta, H.-W. Lin and B. Yoon, Phys. Rev. Lett. {\bf 115}, 212002 (2015).

\bibitem{pndmetensor2}
T. Bhattacharya, V. Cirigliano, S. D. Cohen, R. Gupta, A. Joseph, H.-W. Lin and B. Yoon, Phys. Rev. D {\bf 92}, 094511 (2015).

\bibitem{weinbergop}
S. Weinberg, 
Phys. Rev. Lett. {\bf 63} (1989) 2333.

\bibitem{choi}
K. Choi, S. H. Im, H. Kim and D. Y. Mo, 
Phys. Lett. B {\bf 760}, 666 (2016). 

\bibitem{pospelovweinbergop}
D. Demir, M. Pospelov and A. Ritz, 
Phys. Rev. D {\bf 67}, 015007 (2003). 

\bibitem{pospelovcpvgpinn}
M. Pospelov, 
Phys. Lett. B {\bf 530}, 123 (2002). 

\bibitem{gudkovreview}
V. P. Gudkov, 
Phys. Rep. {\bf 212}, 77 (1992).

\bibitem{bunakov}
V. E. Bunakov and V. P. Gudkov, 
Nucl. Phys. A {\bf 401}, 93 (1983).

\bibitem{bowman}
J. D. Bowman and V. Gudkov, 
Phys. Rev. C {\bf 90}, 065503 (2015).

\bibitem{ibrahim}
T. Ibrahim and P. Nath, 
Phys. Rev. D {\bf 58}. 111301 (1998) [Erratum ibid. D {\bf 60}, 099902 (1999)]. 

\bibitem{ramseyli}
Y. Li, S. Profumo and M. Ramsey-Musolf, 
JHEP {\bf 1008}, 062 (2010). 

\bibitem{ellisgeometric}
J. Ellis, J. S. Lee and A. Pilaftsis, 
JHEP {\bf 1010}, 049 (2010).

\bibitem{ellisgeometric2}
J. Ellis, J. S. Lee and A. Pilaftsis, 
JHEP {\bf 1102}, 045 (2011). 

\bibitem{rpvlinearprogramming}
N. Yamanaka, T. Sato and T. Kubota, 
JHEP {\bf 1412}, 110 (2014). 

\bibitem{sargsian}
M. M. Sargsian, chapter to be published in ``$NN$ and 3$N$ Interactions'' (NOVA Science Publishers, Inc.) [arXiv:1403.0678].

\bibitem{miller}
G. A. Miller, M. D. Sievert and R. Venugopalan, Phys. Rev. C {\bf 93}, 045202 (2016).

\bibitem{hal1}
N. Ishii, S. Aoki and T. Hatsuda, 
Phys. Rev. Lett. {\bf 99}, 022001 (2007).

\bibitem{hal2}
S. Aoki, T. Hatsuda and N. Ishii, 
Prog. Theor. Phys. {\bf 123}, 89 (2010).

\bibitem{sakuda1}
T. Sakuda and F. Nemoto, Prog. Theor. Phys. {\bf 62}, 1274 (1979).

\bibitem{sakuda2}
T. Sakuda and F. Nemoto, Prog. Theor. Phys. {\bf 62}, 1606 (1979).

\bibitem{inoue3}
S. Inoue, M. J. Ramsey-Musolf and Y. Zhang, Phys. Rev. D {\bf 89}, 115023 (2014).

\bibitem{dekens3}
Y.T. Chien, V. Cirigliano, W. Dekens, J. de Vries and E. Mereghetti, JHEP {\bf 1602}, 011 (2016).

\bibitem{fuyuto2}
K. Fuyuto, J. Hisano and E. Senaha, Phys. Lett. B {\bf 755}, 491 (2016).

\bibitem{Kobakhidze}
A. Kobakhidze, N. Liu, L. Wu and J. Yue, Phys. Rev. D {\bf 95}, 015016 (2017). 

\bibitem{Balazs}
C. Balazsa, G. White and J. Yue, JHEP {\bf 1703}, 030 (2017). 

\end{thebibliography}
\end{document}